\documentclass[pra,aps,amsmath,amssymb,superscriptaddress,longbibliography, notitlepage,footinbib]{revtex4-1}

\usepackage{graphicx}
\usepackage{color,outlines}
\usepackage{hhline}
\usepackage{physics}
\usepackage{adjustbox}
\usepackage{hyperref}
\usepackage{amsthm}
\theoremstyle{plain}
\usepackage{amsfonts}
\usepackage{cleveref}
\usepackage{physics}

\newcommand{\AB}{\mathcal{A}\prec \mathcal{B}}
\newcommand{\BA}{\mathcal{B}\prec \mathcal{A}}

\newtheorem{thm}{Theorem}

\newtheorem{corollary}{Corollary}
\newtheorem*{proof*}{Proof}

\makeatletter
\newcommand\footnoteref[1]{\protected@xdef\@thefnmark{\ref{#1}}\@footnotemark}
\makeatother
\begin{document}

\title{Advantage in two-way communication using non-classical states of light} 
\author{Bohnishikha Ghosh}
\email{bg14ms043@iiserkol.ac.in}
\affiliation{Department of Physical Sciences, IISER Kolkata, Mohanpur 741246, India}
\author{Amit Mukherjee}
\email{amitisiphys@gmail.com}
\affiliation{Optics     and    Quantum
  Information Group, The Institute of Mathematical Sciences, HBNI, CIT
  Campus,  Taramani,  Chennai  600113, India}
 \author{S. Aravinda} 
 \email{aravinda@physics.iitm.ac.in}    \affiliation{Department of Physics, Indian Institute of Technology Madras, Chennai 600036, India}



\begin{abstract}
The advantage of using a single-photon two-mode entangled state in two-way communication via maximal violation of an inequality associated with the `Guess Your Neighbour's Input' (GYNI) game has been theoretically [Phys. Rev. Lett. {\bfseries 120}, 060503 (2018)] as well as experimentally [CLEO {\bfseries FID.4} (OSA, 2018)] established quite recently. We show that such an advantage can also be obtained using any single-mode pure non-classical state embedded in a two-mode pure entangled state, wherein the other mode is the vacuum (henceforth referred to as a generalized NOON state), regardless of the average photon number of the single mode state. For the special cases of the even-coherent, odd-coherent, and squeezed vacuum NOON states, we establish that the advantage is also maximal. We show that the usage of the even-coherent NOON states can provide an advantage under noisy apparatuses (beam splitters and photo detectors). As an aside, we study how some of these states fare in terms of violation of a reference-frame independent Bell-type inequality. 
\end{abstract}

\maketitle
%%%%%%%%%%%%%%%%%%%%%%%%%%  body  %%%%%%%%%%%%%%%%%%%%%%%%%%

\section{Introduction}
\label{sec:intro}
 The genesis of quantum mechanics (QM) marks the birth of the distinction between our understanding of the classical world and that of the non-classical world. This aspect took center stage during sixties after the introduction of quasi-probability P-function in characterizing non-classical states of light \cite{Sud63,GR63} and the shortfall of deterministic local hidden variable models \cite{EPR35,Bel64} (and of non-contextual deterministic hidden variable models \cite{KS67}) in explaining the predictions of QM. These concepts contribute to the keystone of our understanding of the difference between the classical and quantum descriptions of the natural world. These fundamental aspects were addressed while making use of QM to obtain technological leaps in communication and computation. In fact, the advent of quantum entanglement, non-locality \cite{EPR35,Bel64} and quantum optical non-classical states \cite{MW95,Lou00}, helped to provide us an enormous advantage compared to classical theories in tasks involving communication and computation \cite{BCP:14,BCM+10,Vei+12}. In this regard, recent developments \cite{DD18,Mas+18} report that two-way communication between two distant parties using a single information carrier can be achieved with the help of superposition of states. In the present work, we delve into the possibility of achieving two-way communication using such single quantum systems which are prepared as non-classical states of light.

Einstein, Podolsky and Rosen (EPR) \cite{EPR35} were the first to question the possibility of explaining the quantum mechanical predictions by local realistic description. Bell retorted by proving the impossibility of explaining certain predictions of QM by any local realistic model \cite{Bel64}, thus establishing the failure of the classical description. He introduced the set of constraints on the observed statistics, named after him as Bell-type inequalities, that any local realistic hidden variable (HV) model should satisfy. The violation of Bell-type inequalities thus characterize the non-classical correlations. %Many loophole free experiments \cite{FC72,ADR82,SMC+15,HBD+15}  were performed to confirm the non-local nature of our physical world. 
Other methods of separating classical and quantum descriptions with the aid of concepts like contextuality \cite{KS67}, discord \cite{OZ01}, non-simplicial structure of state space \cite{ASP17} etc  also exist. %Here it is important to note that the relation among various notions of such non-classical aspects have been studied widely (e.g. \cite{JH14, OW10,BGG+0}). 
These types of non-classicalities can be broadly characterized as quantum information theoretic non-classicality.

In close juxtaposition, the concept of classicality in quantum optics is defined as the viability of expressing any quantum state as a convex combination of coherent states \cite{Sud63,GR63}\footnote{It must be noted here that the only single-mode pure states which are classical, in the sense of quantum optics, are the coherent state.}. In general this is achieved by expressing any state in terms of quasi-probability distribution functions, for example Sudarshan-Glauber P-distribution function \cite{Sud63,GR63}. The negativity and/or singularity of the P function indicates the non-classicality of quantum states \cite{Man86,Vog00}. 
%Loud80,Mir+10,RV02,
 
 Despite great efforts exerted towards understanding the non-classicalities of quantum world in their individual arenas, we lack the quantitative relation between these two notions, with a few exceptions \cite{FP12,BBO+15,Spek08cont}. In the light of technological developments, it is important to study the interplay between the two aforementioned types of non-classicality, particularly by focusing on the advantage gained in information theoretic task  (for example, quantum communication) on using non-classical states of light. In this work, we consider this as a prime motivation to understand the relation between an optical notion of non-classicality and its relation to an information theoretic task.

%GHZ95,Vaid95,TWC91
In most cases, non-locality and entanglement are attributed to
composite systems involving multiple particles as opposed to single particle systems. The erstwhile controversy (\cite{Hard95,DV07} and references therein) on single photon entanglement and non-locality has now been resolved \cite{Enk05,BCB13} and this, as a resource, has been successfully used for performing communication tasks \cite{SS+11,SZ12}. Recently, by exploiting the superposition of a single quantum particle in two spatially separated distant locations, a communication advantage was exhibited, in which the protocol is restricted by usage of the single particle's (or a single information carrier) finite speed of propagation \cite{DD18,Mas+18}. Del Santo \textit{et al.}, via their gedanken experiment, achieved \textit{two-way} communication, i.e, simultaneous communication (within the given time window) of information between both the parties involved, a task that is impossible to perform classically. This is shown by furnishing violation of an inequality associated with a causal game called the `Guess Your Neighbour's Input' (GYNI) game \cite{Cyr16}.
Massa \textit{et al.} experimentally verified the aforementioned proposal.

The common theme in both scenarios of non-locality and the two-way communication is the usage of the single photon entangled state,
 \begin{equation}
	 \ket{\Psi}_{AB} = \frac{1}{\sqrt{2}} (\ket{1}_A \ket{0}_B + \ket{0}_A \ket{1}_B),
	 \label{eq:sing}
 \end{equation}
 where $\ket{i}_{A,B}$ denotes the occupation number state of mode A, B, respectively. From a practical perspective, it is well known that single photons are difficult to produce deterministically and sources which use parametric down-conversion for the production of single photons, have inherent drawbacks. %\cite{LHA+01,HOM87,PJF05,HAK11,WRF+08,ASW06,MAH+09,OCL+08,SSB+12} 
 The trade off between these drawbacks have been investigated by Christ and Silberhorn (refer to \cite{Silb12} and references therein). Based on this practical difficulty, we have been prompted to ask whether the advantage in two-way communication is restricted to the usage of single-photon entangled states, thus initiating a generalization of the existing protocol, invoking quantum optical non-classicality for the proposed task.
 
 In the present work, we explore the possibility of the quantum optical non-classical state of the form 
 \begin{equation}
  \ket{\Phi}_{AB} = \frac{1}{\sqrt{\mathcal{N}}} (\ket{\xi}_A \ket{0}_B + \ket{0}_A \ket{\xi}_B),
  \label{eq:NOON}
 \end{equation}
 where $ \ket{\xi}$ is a single-mode non-classical state, $\mathcal{N}$ is the normalization factor, furnishing a violation of a GYNI inequality \cite{gen_r} (and of a Bell-type inequality). We establish that non-classicality of the single-mode pure state $\ket{\xi}$ is a necessary and sufficient condition for $\ket{\Phi}$ to furnish a violation of a GYNI inequality. We show that the spatially separated superpositions of even (odd)-coherent \cite{Ger93,VBK92} and squeezed \cite{Dod02} states, called the even (odd)-coherent NOON state and the squeezed NOON state respectively, can supply as much advantage as the spatially superposed single photon state, in the task of two-way communication. This result is interesting in the light of recent studies that link the quantum optical non-classicality to quantum computational advantages over classical computation \cite{AA11, RTC16,Vei+12,Vei13}, and in metrology \cite{KK+19}. As $\ket{\Phi}_{AB}$ is a two-mode entangled state involving a single-mode non-classicality, we shall refer to it as the `generalized NOON state', drawing inspiration from the NOON states \cite{Dowling02}. At this stage, it would also be prudent to make a resource-wise comparison between two-way communication and non-locality, given the difference in these two tasks and their associated polytopic structures.  We discuss whether or not being non-local (of a two-mode pure entangled state) has any bearing on its usefulness in the task of two-way communication. 

 Section \ref{sec:prereq} is aimed at helping the reader to be acquainted with some of the prerequisites. This is followed by a descriptions of the protocol of the task of two-way communication and its execution using specific states \ref{sec:two_com}. In section \ref{subsec:ncl} we delve into the significance of single-mode non-classicality in the aforementioned task. In the section that follows (section \ref{sec:loss}), we discuss models of loss in the apparatuses used in the protocol, and make a comparison between the even-coherent NOON state and the single-photon entangled state. We analyze the non-locality of some of the states being studied, in section \ref{sec:nonloc}. The final section \ref{sec:conc} summarizes our results and is indicative of possible future directions.

\section{Prerequisites \label{sec:prereq}}
In this section, we attempt to elucidate the premise for two-way communication as well as the exhibition of non-locality. 

At the outset, let us consider the task of two-way communication. In the  simplest scenario, two parties Alice and Bob (say), are separated by a spatial distance $d$, and have to communicate their messages to each other. The speed of message transmission is restricted by the fact that a single information carrier can traverse the distance between the two parties only once within the given time window. The communication task is formulated as a game called GYNI, in which a referee provides inputs to the players Alice and Bob, and each player has to predict other player's input via his/her output process. 
 Suppose, both Alice and Bob have a two-input-two-output device, and  Alice (Bob) inserts her (his) input $x (y)$ and obtains an output $a (b)$. The task of each player is to predict other player's input, using a single information carrier, within the time interval ($\tau \leq \frac{d}{c}$), taken by a single information carrier to travel the distance $d$ between the two parties.
 
Within the classical model, Alice (Bob) can encode her (his) message in 
the particle and send it to Bob (Alice). If Alice's action precedes Bob's, Alice can send (signal) her message to Bob but Bob can't send his message to Alice. In a causal structure, this situation is denoted by $\AB$ (Alice's actions precedes Bob's) and marginal probability of Alice's outcome is unaffected by Bob's action: $P (a|x,y) =  P (a|x,y^\prime),$ where $  P (a|x,y) = \sum_b P (a,b|x,y)$ and $  P (a|x,y^\prime) = \sum_b P(a,b|x,y^\prime)$. Analogously, we can denote a causal structure in which  Bob's action precedes Alice's ones, by $\BA$, and we have $	P (b|x,y) =  P (b|x^\prime,y)$.

The set of all  correlations $P(a,b|x,y)$, such that $P(a,b|x,y) \geq 0 $ and $\sum_{a,b}P(a,b|x,y) = 1 $  forms a polytope in 12 dimensional vector space, called  \textit{correlation} polytope. In the classical scenario,  the correlations either belong to causal order $\AB$ or to $\BA$ or to any convex combination of these two, thereby forming the \textit{one-way signaling} polytope.
%The set of these correlations are called one-way signaling correlations and it forms \textit{one-way signaling} polytope.%
The facets (represented by some inequalities) of the one-way signaling polytope  give us the maximally achievable classical bound on the probability of success in the GYNI game. In general, for two-party two-input-two-output case, there are two inequivalent sets 
of inequalities called \textit{`Guess Your Neighbor's Input'} (GYNI) and \textit{`Lazy Guess Your Neighbor's Input'} (LGYNI) \cite{Cyr16}. In the present work we restrict ourselves to the \textit{GYNI} inequalities, the set of all 16 of which are merely relabellings of each other. We have worked predominantly with one of these 16 inequalities, and denote it by $\mathcal{J}$,
%\begin{widetext}
\begin{equation}
 \mathcal{J} {\equiv} \frac{1}{4} \left[P(0,0|0,0)+P(0,0|1,1)+P(1,1|0,1)+P(1,1|1,0)\right] \leq \frac{1}{2}.
    \label{eq:gyni}
\end{equation}
%\end{widetext}
Thus, violation of the GYNI inequality (\ref{eq:gyni}) guarantees two-way communication within the time-window $\tau$.

Consider now, the nonlocality scenario involving two parties Alice and Bob, with each of the parties having an access to a device having two inputs and two outputs. In this case, the correlations has to satisfy the no-signaling conditions $P(a|xy)  = P(a|xy^\prime),\hspace{1mm}P(b|xy)  = P(b|x^\prime y)$. These correlations form a no-signaling (NS) polytope in 8 dimensions with 8  nonlocal vertices and 16 local deterministic vertices. The facets of the local polytope are the Clauser-Horne-Shimony-Holt (CHSH) \cite{CHSH} inequalities.

%The point of difference between the geometric structure of the correlation polytope and the no-signaling polytope is worth considering.
The NS polytope is a subset of the correlation polytope and their dimensions are unequal, thus making them two different geometrical entities. This geometric consideration, as well as the different physical scenarios involved in the two-way communication and non-locality tasks, make it nontrivial to compare these two.

\section{Two-way communication}
\label{sec:two_com}
We now describe the method of using some generalized NOON states for carrying out the task of two-way communication. Although, we focus on generalized NOON states involving even (odd)-coherent states, in appendix (\ref{ap:sq}) we show that generalized NOON states involving single-mode squeezed states also violates inequality (\ref{eq:gyni}) maximally.

 \subsection{The protocol \label{sec:proto}}
 At the outset, it is necessary to elucidate the two-way communication protocol described in \cite{DD18} and \cite{Mas+18}. \textbf{(1) Preparation:} Alice and Bob receive a state $\ket{\xi}$ prepared in the superposition of their locations, i.e a generalized NOON state of the form $\ket{\Phi}_{AB}$ (c.f. eq. \ref{eq:NOON}).
  Alice receives mode A and Bob receives mode B of the aforementioned state.  \textbf{(2) Encoding:} Alice and Bob encode their respective inputs as phases on their local states by operating on the polarization degree of freedom. The inputs are the bits $x,y{\in}\{0,1\}$. Naturally, this doesn't work when Alice or Bob receive the vacuum state $|0\rangle$. \textbf{(3) Beam splitter operation:} The encoded even-coherent NOON state is made to pass through a 50:50 beam splitter (BS). In the current section, we assume that the beam splitter is loss-less. A possible treatment of lossy beam splitters is provided in section \ref{sec:loss}. \textbf{(4) Detection:} The outputs of the BS are detected in the local laboratories of Alice and Bob by performing suitable  dichotomic projective measurements. They assign values to their respective output bits $a$ and $b$ ($\in \{0,1\}$) depending on the measurement that clicks. Within the time $\tau$, the two parties must simultaneously obtain their respective outputs.

\subsection{Execution of the protocol for the even-coherent NOON state}
\label{sec:exec}
The protocol described in section \ref{sec:proto} can be carried out as described below.
An even-coherent NOON state is of the form:
  \begin{equation}
      \ket{\Phi_{E}} = \frac{1}{\sqrt{\mathcal{N}(\alpha)}}[\ket{\alpha_e}\ket{0}+\ket{0}\ket{\alpha_e}],
      \label{eq:evco}
  \end{equation}
  where $\ket{\alpha_e} \propto [\ket{\alpha}+\ket{-\alpha}]$, i.e a Cat state \cite{Ger93,VBK92} ($\ket{\alpha}$ is a coherent state with amplitude $\alpha$). The subscripts for the modes have been omitted for the sake of simplicity. Here, $\mathcal{N}(\alpha) = \frac{1}{\sqrt{2 [1+\exp(-2|\alpha|^2)]}}$ is the normalization constant
   Although we have concentrated on even-coherent NOON states, it is to be noted that the protocol remains valid for odd-coherent NOON states as well.
 In the number basis, $\ket{\alpha_e}$ is represented as 

\begin{equation}
 \ket{\alpha_e} = \frac{1}{\cosh{|\alpha|^2}} \sum_{n=0}^\infty \frac{\alpha^{2n}}{\sqrt{(2n)!}}{\ket{2n}}
\end{equation}

\begin{figure}[ht]
\centering
	\includegraphics[width=7cm, height=6cm]{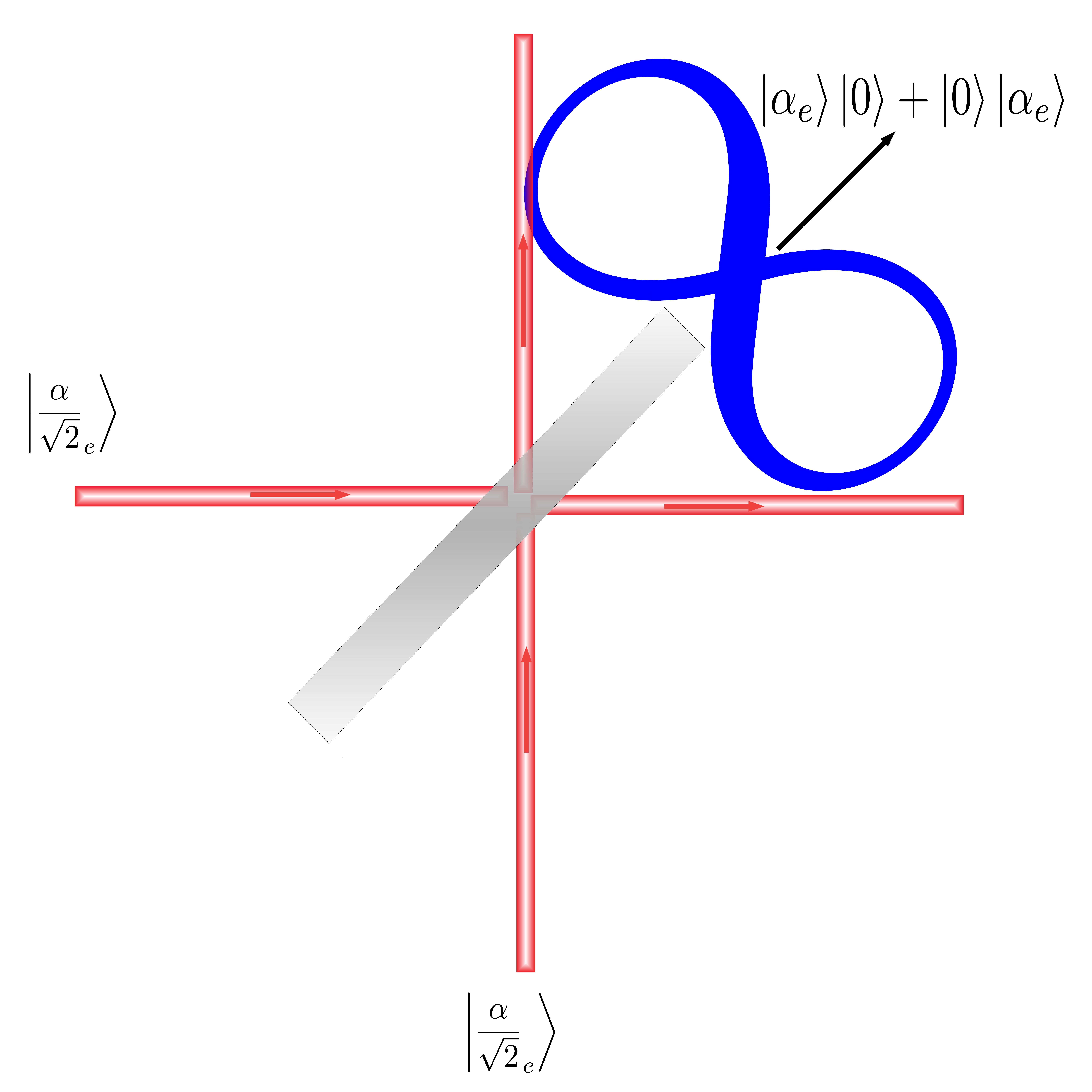}
	\caption{(Colour online) The state $\frac{1}{\sqrt{\mathcal{N}(\alpha_e)}}[\ket{\alpha_e}\ket{0}+\ket{0}\ket{\alpha_e}]$, i.e the even coherent NOON state, can be prepared by using the state $\ket{\frac{\alpha}{\sqrt{2}}_e}$ in each of the BS. Note that the normalization factor of the state $\ket{\Phi_E}$ has been omitted in the figure.}
	\label{fig:prep1}
\end{figure}

Studies on even and odd ($\ket{\alpha_o} \propto [\ket{\alpha}-\ket{\alpha}]$) coherent states and their respective preparations have been performed and are still under way. Experimental methods of preparing even-coherent states include, for example, remote state preparation using hybrid entangled states \cite{Jean2018,Morin2014} and measurement on the two level atomic system (this can be performed with high efficiency) of atom-light entangled states \cite{Hacker2019}. Let $\ket{\Phi_E}$ (c.f. eq. (\ref{eq:evco})) be the state shared by Alice and Bob. Here the first mode is with Alice while second mode is with Bob. We describe below the production of the state $\ket{\Phi_E}$ using a 50:50 beam splitter (BS). Let us define the action of 50:50 BS on input modes as $ \hat{a}^\dagger \rightarrow \frac{\hat{a}^\dagger + \hat{b}^\dagger}{\sqrt{2}} \quad , \quad \hat{b}^\dagger \rightarrow \frac{\hat{a}^\dagger - \hat{b}^\dagger}{\sqrt{2}}. $ By considering the action of the loss-less 50:50 BS on the two-mode product states-- $\ket{\alpha}\ket{0}, \ket{0}\ket{\alpha}, \ket{-\alpha}\ket{0}$, and $\ket{0}\ket{-\alpha}$, and by taking into account the fact that the BS is a unitary operation that is its own inverse, we can create even-coherent NOON states. This is summarized in Fig. (\ref{fig:prep1}). 

Alice and Bob encode their respective inputs on each mode of $\ket{\Phi_E}$, in the form of the bits $x,y{\in}\{0,1\}$, into the phase of polarization degree of freedom of their respective modes resulting in the following state, which is represented in number basis upon normalization. 
%\begin{widetext}
\begin{equation}
 \ket{\Phi^{\prime}_{E,xy}} = \frac{e^{\frac{-|\alpha|^2}{2}}}{1+(-1)^{x+y}e^{-|\alpha|^2}} \sum_{n=0}^\infty \frac{\alpha^{2n}}{\sqrt{(2n)!}} [(-1)^x \ket{2n,0} + (-1)^y ]\ket{0,2n}].
 \label{eq:in_state}
\end{equation}
%\end{widetext}
The average photon number of the input state (c.f appendix (\ref{ap:avg})) $\ket{\Phi_E}$ can be tuned to a desirable value, and in principle can be made as small as possible. This number is, in fact, a scalar multiple of the average photon number of the embedded single mode state, i.e, the even-coherent state. 
After encoding their inputs,  they send the encoded state to a 50:50 BS. The action of the BS on $\ket{\Phi^\prime_{E,xy}}$ results in the following state 

 \begin{equation}
  \ket{\Phi^{\prime{\prime}}_{E,xy}}= \frac{e^{\frac{-|\alpha|^2}{2}}}{1+(-1)^{x+y}e^{-|\alpha|^2}} \sum_{n=0}^\infty \frac{\alpha^{2n}}{\sqrt{(2n)!}} \frac{1}{2^n} \sum_{k=0}^{2n} \left(\begin{array}{c} 2n \\ k \end{array}\right)^{\frac{1}{2}} [(-1)^x + (-1)^{y+2n-k}]\ket{k,2n-k}.
  \label{eq:ev_co_out}
\end{equation}

It is evident from eq. ($\ref{eq:ev_co_out}$) that when $(x+y)\hspace{1mm} \text{mod}{\hspace{1mm}} 2 {\equiv}x{\oplus}y=0$ ($x{\oplus}y=1$), Alice and Bob have even (odd) Fock states at their outputs. 
Thus, both of them separately make use of photon number parity resolving projectors: 
\begin{equation}
 P_{even} = \sum_{l=0}^\infty \ketbra{2l}{2l} \quad , \quad P_{odd} = \sum_{l=0}^\infty \ketbra{2l+1}{2l+1}.
	\label{eq:meas}
\end{equation}
The success of the protocol is based on the simultaneous detection within the time window $\tau$ of odd or even number of photons at both Alice and Bob's ends. In the present case, we assume that the detectors at the two ends are perfect. This assumption is relaxed in section \ref{sec:loss}.
The outputs $a,b = 0$ correspond to the clicking of $P_{even}$ at both Alice and Bob's lab. Similarly the outputs $a,b = 1$ correspond to the clicking of $P_{odd}$ at both Alice and Bob's lab. It is clear from eq. (\ref{eq:ev_co_out}), that $P(0,0|0,0) = P(0,0|1,1) = P(1,1|0,1) = P(1,1|1,0) = 1$ and $P(a,b|a,y)$ for all other values of $a,b,x,$ and $y$ vanish. The outcomes $a,b$ satisfy the functional relation $a = x\oplus y$ and $b = x\oplus y$ with $x,y$ being the inputs, thus facilitating Alice and Bob to deterministically predict the input of the other party, consequently furnishing a maximal violation of inequality (\ref{eq:gyni}).

\section{How significant is non-classicality?}
\label{subsec:ncl}
Consider now the coherent NOON state $\ket{\Phi_C}=\frac{1}{\sqrt{\mathcal{N^{\prime}}(\alpha)}}[\ket{\alpha}\ket{0}+\ket{0}\ket{\alpha}]$, where $\mathcal{N^{\prime}}(\alpha)$ is the normalization factor. The very same protocol (c.f. \ref{sec:proto}) can be used to show that this state doesn't furnish any violation of inequality (\ref{eq:gyni}) by noting that this state, on passing through the BS is separable \cite{Gerry2004} and hence there is no way to distinguish the case in which the inputs are correlated from that in which they are anti-correlated. This result illustrates that the non-classicality of the single-mode state $\ket{\xi}$ state (c.f eq. (\ref{eq:NOON})) is certainly a necessary condition for the resultant generalized NOON $\ket{\Phi}$ state to furnish the violation of the GYNI inequality. We have observed that even (odd)-coherent, and squeezed vacuum generalized NOON states furnish maximum violation of the GYNI inequality. In each of these states, the embedded single-mode states are superpositions of either even or odd Fock states. In fact, this is true for any single-mode pure non-classical state $\ket{\xi}$, which is a finite/infinte superpositions of only even (odd) Fock states. Is it possible to find other single-mode non-classical states which are more general superpositions of the Fock states, such that the corresponding generalized NOON states violate (not necessarily maximally) the GYNI inequality? In this regard, we present the following theorem.

\begin{thm}
\label{thm1}
The non-classicality of the single-mode pure state embedded in the genralized NOON state $\ket{\Phi}$ (eq. (\ref{eq:NOON})), is a necessary and sufficient for $\ket{\Phi}$ to furnish a violation of the GYNI inequality of eq. (\ref{eq:gyni}).
\end{thm}
\textbf{Proof:}
Suppose $\ket{\xi}=\sum^L_{n=0}\lambda_n\ket{n}$ ($\lambda_n{\in}\mathbb{C};{\hspace{1mm}} \sum^{L}_{n=0}|\lambda_0|^2=1$), i.e., a single-mode normalized pure state. $L$ maybe finite or infinite. If $L$ is finite, we know via \cite{Park2017}, that $\ket{\xi}$ is a non-classical state, provided that $\lambda_0{neq}1$. We also know that when $L$ is infinite, $\ket{\xi}$ is classical, if and only if it is a coherent state (of some amplitude). Consider the generalized NOON state defined in eq. (\ref{eq:NOON}). The theorem states that the state $\ket{\Phi}$ furnishes a violation of a GYNI inequality, if and only if $\ket{\xi}$ is non-classical. A few steps of algebra can show that on Bob's side, the measurement to be chosen in order to distinguish between the outputs based on the parity of the inputs is the dichotomic projective measurement $\{P_{even},P_{odd}\}$, given in eq. (\ref{eq:meas}). One can then establish that irrespective of whether $L$ is finite or infinite, a suitable set of dichotomic projective measurements can be chosen on Alice's side in order to furnish a violation of the GYNI inequality. The choice of the set of projectors on Alice's side can be optimized to furnish maximum possible violation for a given state, which is not necessarily the maximum value that the quantity $\mathcal{J}$ can take. The reader is requested to refer to appendix (\ref{subsec:proof}) for the details.

\section{Lossy Apparatus}
\label{sec:loss}
We now describe some models of loss in the devices used in the aforementioned protocol. 
\subsection{Lossy beam splitters}
\label{subsec:bs_loss}
Suppose $\mathbf{T}(\omega)$ is the is the $2\times2$ unitary matrix, corresponding to a lossless beam splitter with transmittivity $t(\omega)$ and reflectivity $r(\omega)$ satisfying $|t(\omega)|^2+|r(\omega)|^2=1$ and $t(\omega)r^{*}(\omega)+r(\omega)t^{*}(\omega)=0$, for all angular frequencies $\omega$. 
 The imaginary part the dielectric permittivity of matter (that constitutes the beam splitter) contributes to the losses in the system. In presence of losses (absorption of radiation) in the system, the above-described unitary relation doesn't hold and $|t(\omega)|^2+|r(\omega)|^2\leq1$, with the equality holding in the lossless case only \cite{Barnett98}. The sources of loss must be taken into account so as to preserve the canonical commutation relations between the outgoing field's mode operators.

\begin{figure}[ht]
    \centering
    \includegraphics[width=5cm, height=4cm]{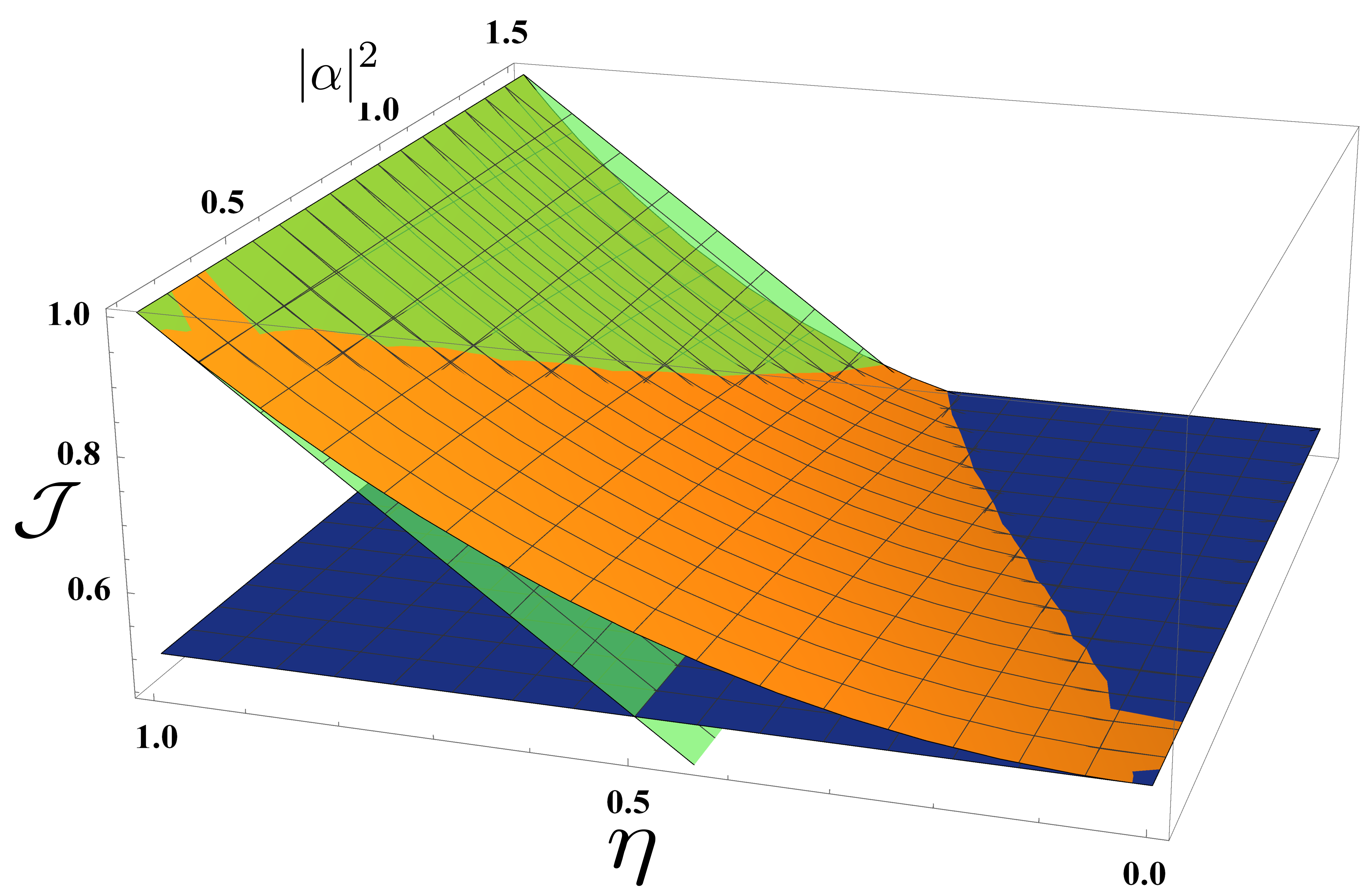}
    \caption{(Colour online) The blue plane marks the classical bound of the GYNI inequality. The green plane in the plane that gives the LHS of the GYNI inequality $\mathcal{J}$ for the single photon state. The surface in yellow is the GYNI inequality given by the even-coherent NOON state, plotted as a function of $\eta$ and $|\alpha|^2$. It is to be noted that the yellow curve shows violation of the GYNI inequality for quantum efficiency $\eta$, of the beam splitter, less than $0.5$ and small values of $|\alpha|^2$ and hence of the average photon number.}
    \label{fig:2}
\end{figure}

  A Kramers-Kronig consistent quantization scheme of the electromagnetic field in dispersive and absorbing inhomogeneous media has been provided in ref. \cite{GRN96}. This work, has in turn been used to derive the unitary transformation that relates the output quantum state to the input quantum state by Kn{\"o}ll \emph{et. al} \cite{KND98}. For the sake of completeness, we shall briefly describe this formalism. The idea is to define a U(4) matrix such that it transforms four input modes, two of the incoming field and two of the device, to four output modes, of which two are of the outgoing field and the other two are of the device. This approach thus allows for mode and energy conservation of the entire system whose constituents are the field and the device. Mathematically, this action may be represented as follows.
  \begin{equation}
    \boldsymbol{\hat{\beta}}(\omega) = \mathbf{\Lambda}(\omega)\boldsymbol{\hat{\alpha}}(\omega),
\end{equation}
where $\boldsymbol{\hat{\alpha}}(\omega) =
 \bigl(\hat{a}_1(\omega),           
 \hat{a}_2(\omega),
 \hat{g}_1(\omega),
 \hat{g}_2(\omega)\bigr)^T $ and $\hat{a}_j(\omega)$ ($j \in \{1,2\}$) are the amplitude operators of the incoming damped wave at frequency $\omega$, while $\hat{g}_j(\omega)$ describe device excitations by playing the role of the loss generating operators. Analogously $\boldsymbol{\hat{\beta}}(\omega) =
\bigl(\hat{b}_1(\omega),           
 \hat{b}_2(\omega),
 \hat{h}_1(\omega),
 \hat{h}_2(\omega)\bigr)^T$ is the four-dimensional output vector operator and $\hat{b}_j(\omega)$ and $\hat{h}_j(\omega)$ ($j \in \{1,2\}$) are the corresponding output mode operators of the outgoing damped wave and the device, respectively, at frequency $\omega$. As mentioned previously, $\mathbf{\Lambda}(\omega) \in$ U(4), and is given as follows.
\begin{equation}
 \mathbf{\Lambda}(\omega) =    
 \begin{pmatrix}
\mathbf{T}(\omega) & \mathbf{\Lambda}(\omega) \\
-\mathbf{S}(\omega)\mathbf{C}^{-1}(\omega)\mathbf{T}(\omega) & \mathbf{C}(\omega)\mathbf{S}^{-1}(\omega)\mathbf{A}(\omega)
\end{pmatrix},
\label{eq:u4}
\end{equation}
where the matrices $\mathbf{T}(\omega)$ and $\mathbf{A}(\omega)$ are $2\times2$ matrices that represent the transmission and absorption respectively. They satisfy the following relation.
\begin{equation}
    \mathbf{T}(\omega)\mathbf{T}^{\dagger}(\omega)+\mathbf{A}(\omega)\mathbf{A}^{\dagger}(\omega) = \mathbf{I}
\end{equation}
It must also be noted that $\mathbf{C}(\omega)$ and $\mathbf{S}(\omega)$ in equation (\ref{eq:u4}), are commuting positive Hermitian matrices given by the following-- $ \mathbf{C}(\omega) = \sqrt{\mathbf{T}(\omega)\mathbf{T}^{\dagger}(\omega)},\quad\mathbf{S}(\omega) = \sqrt{\mathbf{A}(\omega)\mathbf{A}^{\dagger}(\omega)}.$
Quite evidently, the dependence on the input device modes arise due to the presence of absorption:-- $\boldsymbol{\Hat{b}}(\omega) = \mathbf{T}(\omega)\boldsymbol{\Hat{a}}(\omega) + \mathbf{A}(\omega)\boldsymbol{\Hat{g}}(\omega),$
where $\boldsymbol{\Hat{b}}(\omega), \boldsymbol{\Hat{a}}(\omega)$, and $\boldsymbol{\Hat{g}}(\omega)$ are $2{\times}1$ column vectors of the output field operators, the input field operators, and the input device operators respectively.
For our purpose of describing a lossy beam splitter, we have chosen $\mathbf{T}$ and $\mathbf{A}$ to be frequency independent and of the following form.
\begin{equation}
    \mathbf{T} = \sqrt{\frac{\eta}{2}}
    {\begin{pmatrix}
     1 & 1\\
     1 & -1
    \end{pmatrix}},
    \quad
     \mathbf{A} = \sqrt{\frac{1-\eta}{2}}{
    \begin{pmatrix}
     1 & 1\\
     -1 & 1
    \end{pmatrix}},
     \label{eq:absor}
\end{equation}
where $\eta \in [0,1]$. The reader is requested to refer to appendix (\ref{ap:subsec_bs_extra}) for the form of $\mathbf{\Lambda}$ used in our calculations.
The even-coherent NOON state post-phase encoding and appending with ancillary modes, considering the initial modes of the device to be in the state vacuum, on passing through the lossy form of 50:50 beam splitter, gives the following output state.
\begin{align}
   \ket{\Tilde{\Phi}^{\prime{\prime}}_{E,xy}} = A(x,y,\alpha)\sum_{n=0}^\infty \left(\frac{\alpha}{2}\right)^{2n} \sum_{k_1=0}^{2n} \sum_{k_2=0}^{2n-k_1} \sum_{k_3=0}^{2n-k_1-k_2}\frac{[(-1)^{x+k_1+k_2} + (-1)^{y+k_2+k_3}]}{\sqrt{k_1!k_2!k_3!(2n-k_1-k_2-k_3)!}}\nonumber\\ {\times}\sqrt{\eta^{k_1+k_2}(1-\eta)^{2n-k_1-k_2}}\ket{k_1,k_2,k_3,2n-k_1-k_2-k_3},  
\end{align}
%\end{widetext}
where $A(x,y,\alpha)=\frac{e^{\frac{-|\alpha|^2}{2}}}{1+(-1)^{x+y}e^{-|\alpha|^2}}$. Thus, on tracing over the device modes of $\Tilde{\rho}^{\prime}_{out} = \ketbra{\Tilde{\Phi}^{\prime{\prime}}_{E,xy}}{\Tilde{\Phi}^{\prime{\prime}}_{E,xy}}$ (c.f. appendix (\ref{ap:subsec_bs_extra})) and on performing the projective measurements described in eq. (\ref{eq:meas}), we obtain the probabilities $P(0,0|0,0)=P(0,0|1,1)$ and $P(1,1|0,1)=P(1,1|1,0)$. It is to be noted that the detectors at both ends are assumed to be ideal. If this assumption is relaxed, then corresponding probabilities would be a function of other parameters as well. By the above-described method, the left hand side (LHS) of the GYNI inequality (\ref{eq:gyni}), furnished by the even-coherent NOON state, can be obtained as a function of $\eta$. In order to draw a comparison between the effect of the lossy beam splitter (assuming the aforementioned model) on the phase encoded version of the single photon entangled state (eq. (\ref{eq:sing})), has been studied. The measurements used in this case are consistent with ref. \cite{DD18,Mas+18}, i.e, testing the existence of a single photon at Alice or Bob's end, after the completion of the protocol. It tuns out that the LHS of the GYNI inequality (\ref{eq:gyni}) is given by $\eta$ in this case. These results are summarized in Fig. (\ref{fig:2}). It is noteworthy that the even coherent NOON state furnishes a violation of the chosen GYNI inequality for quantum efficiency $\eta{\leq}0.5$. However, the single photon entangled state shows no violation in this regime. For an alternate description of loss in the beam splitter, the reader is requested to refer to appendix (\ref{ap:subsec_alt}).

%%%%%%%%%%%%%%%%%%%%%%%%%%%%%%%%%%%%%%%%%%%%%%%%%%
\subsection{Loss at the detection end}
{ It is evident from eq. (\ref{eq:ev_co_out}) that for the even-coherent NOON state, Alice and Bob need to have detectors, which can account for several photons simultaneously arriving at the detector, at their disposal. Such detectors are referred to as photon-number-resolving detectors (PNRD) in the existing literature \cite{Eisaman2011,Hadfield2009}. At the detection end, in the absence of loss, it suffices to study the output using projectors on the even and the odd subspace, as given in eq. (\ref{eq:meas}). However, in the presence of loss, projective measurements have to be replaced by positive operator valued measurements (POVMs), taking the factors that contribute to the loss into consideration.
Till date, there exist two different mathematical models of PNRDs, taking loss into account. In the first (Tan \textit{et al.} \cite{Tan16}) model, the photo-counting statistics follow a binomial distribution accounting for losses, and in the second (Sperling \textit{et al.} \cite{Sperling2012,Sperling2012}) model, the incident beam is first split and then each part is measured by on/off detectors. We shall use both models to compare the performances of the even coherent NOON state and the single photon state. The first of these models and its use in the comparative study have been described in appendix (\ref{ap:tan}).

We shall briefly describe and use the Sperling-Vogel-Agarwal model \cite{Sperling2012} here. Suppose we have a $N$ dimensional unitary operator, $U(N)$, on one port of which we impinge a coherent state $\ket{\beta}$. The other $N-1$ ports have vacuum inputs. The unitary operator acts on the coherent state to split it into $N$ modes of equal reduced amplitude (and photon number):--$\ket{\frac{\beta}{\sqrt{N}}}^{{\otimes}N}$. If $N$ is a large number, the average photon number of each of these modes can be approximately unity or less. Now each of these output modes are detected via on/off detectors which can distinguish between the case when no photons are impinged and when some photons are impinged, i.e., they are not photon-number resolving detectors. Examples of such detectors include single-photon avalanche photodiodes and photo-multiplier tubes \cite{Eisaman2011}. We shall assume that $k$ of the $N$ on/off detectors click. This implicitly takes saturation into account. Thus, photo-multiplexing allows us to create a model of PNRDs detectors without having to use PNRDs. In \cite{Sperling2012}, the authors use the above-described idea to define the following POVM.
\begin{equation}
    \Pi_k = :\frac{N!}{k!(N-k)!}e^{-(\kappa\frac{\hat{n}}{N}+\nu)(N-k)}(\hat{I}-e^{-(\kappa\frac{\hat{n}}{N}+\nu)})^k: \hspace{2mm}\forall{k}\hspace{2mm} {\in} \{0,1,2,...,N\},
\end{equation}
where :.: denotes normal ordering (without using commutation relations), $\hat{n}$ is the number operator, $\kappa$ is the quantum efficiency of the system and is typically less than unity, $\nu$ is a measure of dark count (typically ${\geq} 0$).

For our purpose of comparison between the even-coherent NOON state and the single-photon entangled state, we choose the following measurement on Alice and Bob's subsystems, keeping in mind the measurements used when the detectors are perfect.
\begin{equation}
    \mathcal{M} = \{\Pi_0=\sum^N_{k=0}\Pi_{2k},{\hspace{1mm}}\Pi_1=\sum^N_{k=0}\Pi_{2k+1}\}.
\end{equation}
Let us now define the outcomes according to the measurement that clicks, when the input is the even-coherent NOON state:-- $\mathcal{M^{\prime(A)}} = \{a=0 {\equiv} \Pi_0; a=1 {\equiv} \Pi_1\}, \quad
    \mathcal{M^{\prime(B)}} = \{b=0 {\equiv} \Pi_0; b=1 {\equiv} \Pi_1\}$. 
Similarly, the outcomes can be defined when the input is the single-photon entangled state:-- $ \mathcal{M^{\prime{\prime}(A)}} = \{a=0 {\equiv} \Pi_1; a=1 {\equiv} \Pi_0\}, \quad
    \mathcal{M^{\prime{\prime}(B)}} = \{b=0 {\equiv} \Pi_0; b=1 {\equiv} \Pi_1\}$.
Thus, using by making use of these outcome assignments in inequality (\ref{eq:gyni}) we can find the value of the GYNI inequality furnished by the even-coherent NOON state (as a function of $|\alpha|^2,N,\kappa,\nu$) and the single-photon entangled state (as a function of $N,\kappa,\nu$) respectively. We have carried out the comparison in two distinct cases-- when $N$ is even and when it is odd, only to find that not intrinsic difference emerges in these two cases. Hence, we have restricted ourselves to reporting the results for the case in which $N$ is even.

It has been previously mentioned that when $|\alpha|^2{\leq}1$, the average photon number of the even-coherent NOON state is less than or equal to unity. We observe that for a fixed value of $N$, the even-coherent NOON state shows a higher violation than the single-photon state when $\kappa$ is close to 0.5, provided that the dark count $\nu$ is very close to zero. However, this is true only for values of $|\alpha|^2<1$. It is to be noted that the afore-mentioned observation doesn't vary for different values of $N$. These results are summarized in Fig. (\ref{fig:3}). Note that for large values of $N$, the amount of violation furnished by either state exponentially decreases as $\nu$ increases. This is intuitively clear as each on/off detector contributes to the noise and all of their contributions add up when their number is large.
\begin{figure}[ht]
    \centering
    \includegraphics[width=10cm,height=3.7cm]{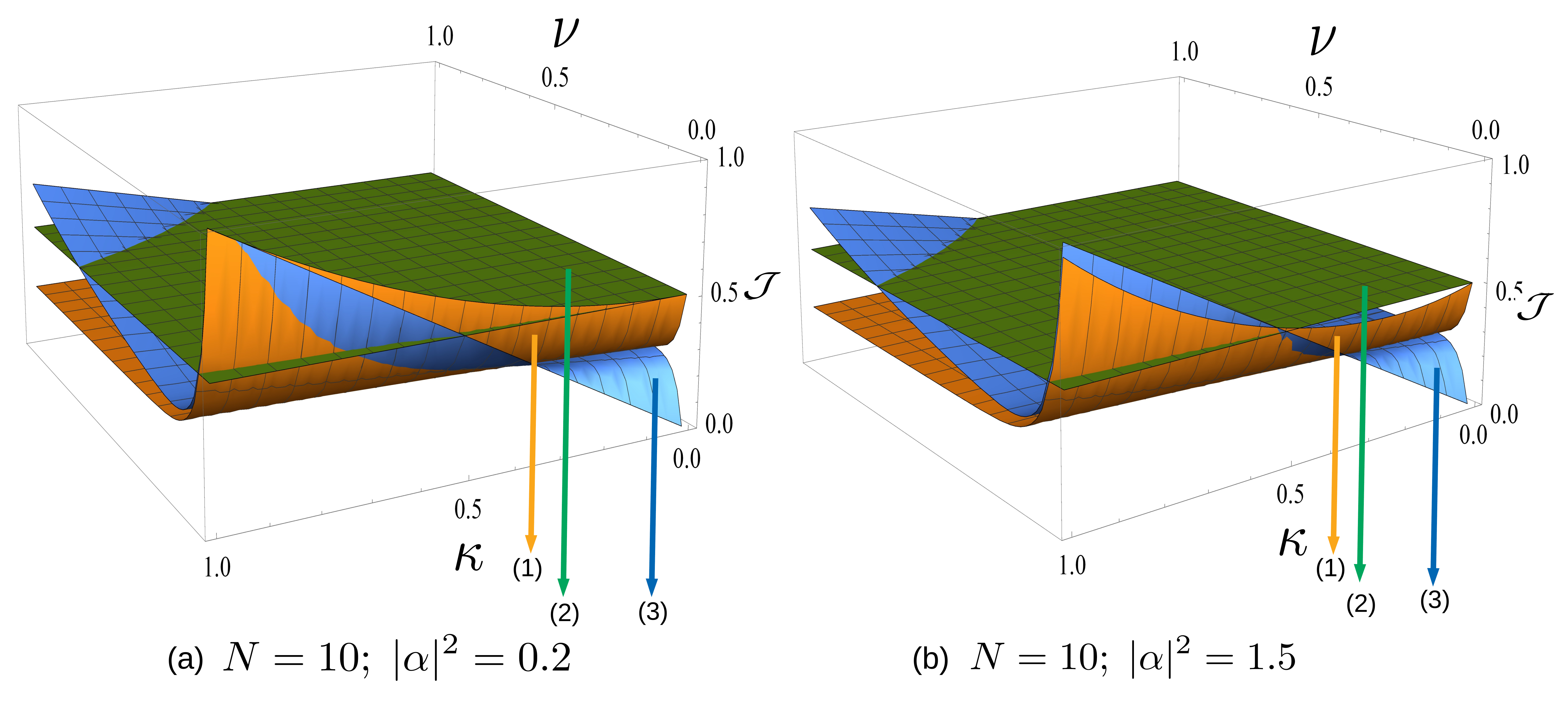}
    \caption{(Colour online) Key: {\bf(1)} threshold value 0.5, i.e classical bound of the GYNI inequality; {\bf(2)} even-coherent NOON state; {\bf(3)} single-photon entangled state. In {\bf(a)}, when $|\alpha|^2=0.2$, note that the even-coherent NOON state furnishes a higher violation of inequality (\ref{eq:gyni}) in comparison to the single-photon entangled state when $\kappa$ is in the neighbourhood of 0.5 and $\nu {\approx} 0$.}
    \label{fig:3}
\end{figure}
}
%%%%nonlocality%%%%%%%%%%%%
\section{Non-locality} 
\label{sec:nonloc}
In this section, as an aside, we perform reference frame independent measurements (c.f. section \ref{sec:intro}) on some the generalized NOON states that we have looked at so far. Entanglement happens to be a necessary but not sufficient condition for non-locality \cite{Werner1989}.  Nevertheless, all bipartite pure entangled states can be shown to be non-local \cite{GISIN91}. Hence there exists at least one measurement scheme, for which the generalized NOON states (which are bipartite pure entangled states) of our concern can be shown to be non-local. However, the measurements that we choose in the following section might not be useful to that end, in every case. Analogously, it is understood that entanglement of the input state is a necessary condition for success in two-way communication but certainly not sufficient, as is shown by the counterexample of the coherent NOON state. Non-classicality of the single-mode, is however, necessary and sufficient. 

 Consider a two party scenario in which both Alice and Bob have two choices of dichotomic measurements $M^A_{0/1}$ and $M^B_{0/1}$, with $\pm 1$ outcomes (for each measurement) respectively. The local realistic correlations satisfy a Bell-type inequality given by \cite{WW01,MB02}
\begin{equation}
	\mathcal{I} = \frac{1}{4} \sum_u |\sum_v (-1)^{u\cdot v} \xi (v)| \leq 1. 
	\label{eq:Belin}
\end{equation}
Here $u,v \in \{0,1\}^2$ and  $\xi (v)\hspace{1mm} (= \expval{M^A_{v_1}M^B_{v_2}})$ is the corresponding correlation function.

\begin{figure}[ht]
\centering
	\includegraphics[width=10cm, height=8cm]{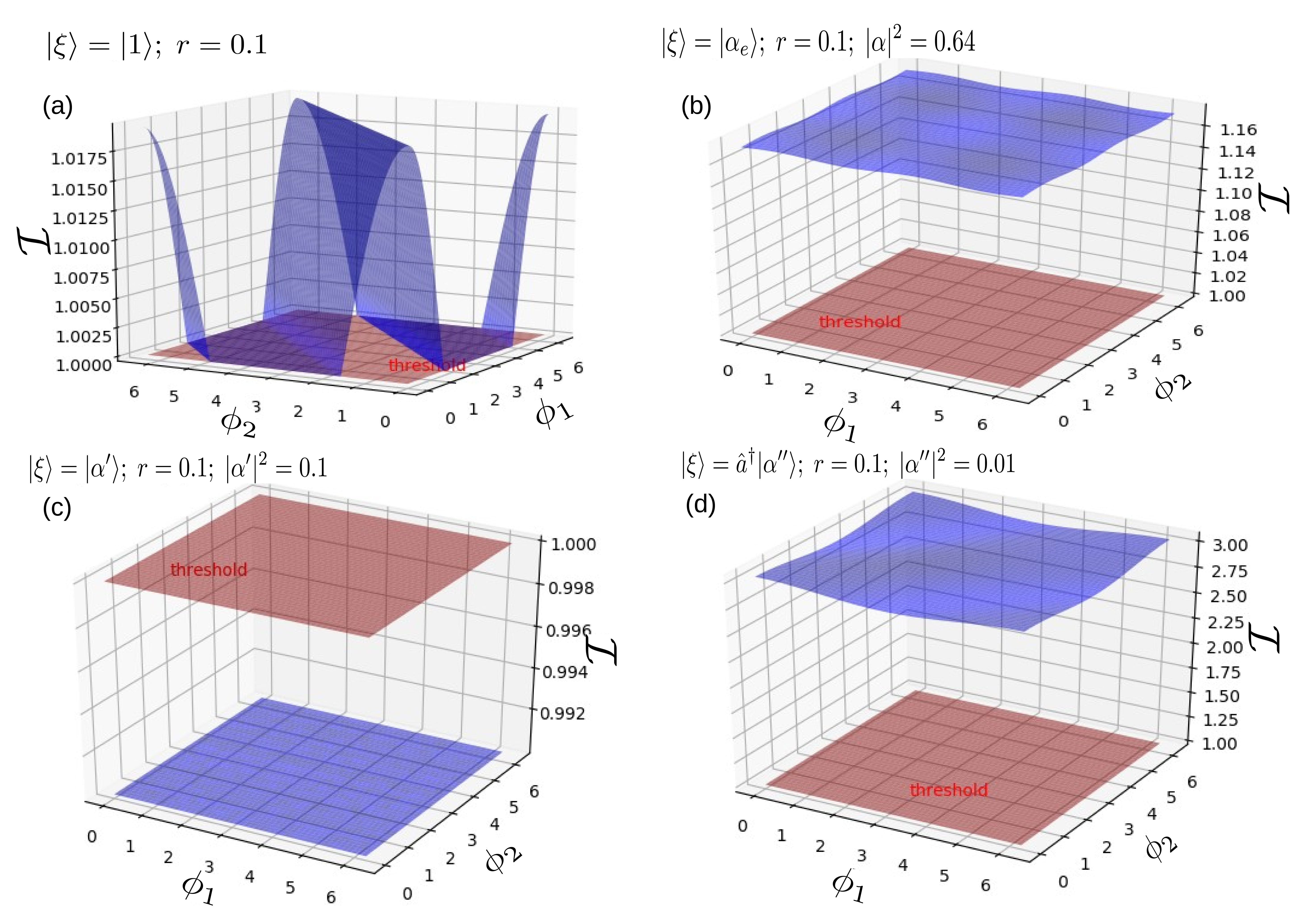}
	\caption{(Colour online) {\bf(a)} The single-photon entangled state (eq. \ref{eq:sing}) shows violation of inequality (\ref{eq:Belin}) for a subset of $\phi_1,\phi_2{\in}[0,2\pi]$. {\bf(b)} The even-coherent NOON state shows a greater violation of inequality (\ref{eq:Belin}) for all values of $\phi_k$ in $[0,2\pi]$. Here $r=0.1$ and $|\alpha|^2=0.64$. {\bf(c)} The coherent NOON state, shows no violation of inequality (\ref{eq:Belin}) irrespective of $\phi_k$s. Here $r=0.1$ and $|\alpha^{\prime}|^2=0.01$. 0.99 is the value closest to the threshold that it can furnish. It can be verified that this holds true for all values of $r$ and $|\alpha^{\prime}|$. {\bf(d)} The photon-added coherent NOON state furnishes a violation for all values of $\phi_k$ and $|\alpha^{\prime{\prime}}|$.}
		\label{fig:1}
\end{figure}

As mentioned in the introduction, the concept of non-locality of a single particle (photon) met with an objection-- the requirement of fixing the reference frame by using a laser beam would lead to the detection of more than one photon and thereby to particle creation. In \cite{BCB13}, the authors, by considering the measurements involving an optical displacement followed by single photon detection, have demonstrated the non-local nature of the single photon state in eq. (\ref{eq:sing}) without the need of a shared reference frame. They have made use of dichotomic measurements of the following form:-- $M^A_0 = M^B_0 = 2\dyad{0} - I$, $ M^A_1 = 2\dyad{\beta_1} - I$ and 
$M^B_1 = 2\dyad{\beta_2} - I$, where $\ket{\beta_k}$ are coherent states. Let us define: $\beta_1 = re^{\jmath \phi_1}$ and $\beta_2 = re^{\jmath \phi_2}$, where $\jmath^2 = -1$ ($r$ is chosen to be the same as this symmetry allows us to maximize the possible violation amount for any state). We make use of the same set of measurements on our states--the even-coherent NOON state, and the coherent NOON state, and the one photon-added coherent NOON state \cite{Park2017} (appendices (\ref{ap:subsec_inf}) and (\ref{ap:subsec_pa})), and compare the results with the violation furnished by the single-photon entangled state. The correlators $\expval{M^A_lM^B_m}, l,m \in \{0,1\}$, for each of the three aforementioned states are given in appendix (\ref{ap:bell}). For $r = 0.1$, inequality (\ref{eq:Belin}) is violated by the even-coherent NOON state for all the values of $\phi_k{\in}[0,2\pi]$ and $|\alpha|$ and the violation is greater than the violation achieved by single photon state $(\ref{eq:sing})$. Similarly, the one photon-added coherent state (with the amplitude of the coherent state being $\alpha^{\prime{\prime}}$) shows a violation of the inequality for all values of $r$ and $|\alpha^{\prime{\prime}}|$. Contrariwise, the state $\frac{1}{\sqrt{\mathcal{N^{\prime}}(\alpha^{\prime})}}[\ket{\alpha^{\prime}}\ket{0}+\ket{0}\ket{\alpha^{\prime}}]$, for all values of $|\alpha^{\prime}|$ and $r$ shows no violation of the aforesaid inequality. These results are summarized in Fig. (\ref{fig:1}). In these figures, the plane plotted in red ochre marks the $z=1$ plane, i.e. the classical bound of inequality (\ref{eq:Belin}). In each of the plots in Fig. (\ref{fig:1}), $\ket{\xi}$ (c.f. eq. (\ref{eq:NOON})) is the single-mode state embedded in the two mode entangled state with the other mode being the vacuum.

\section{Conclusion \& Discussion \label{sec:conc}}

Understanding the relation between optical non-classicality and information theoretic notions of non-classicality enhances our understanding of the quantum world. This understanding will potentially have a significant bearing on technological advancements which make use information theoretic notions of non-classicality. We have proven that all bipartite pure entangled states which have single-mode non-classical states embedded in them, i.e., all generalized NOON states in which $\ket{\xi}$ is non-classical, furnish a violation of the GYNI inequality (\ref{eq:gyni}), irrespective of the average photon number of the state $\ket{\xi}$, thereby exhibiting the advantage in the two-way communication task. We have also established that the even (odd)-coherent and squeezed-vacuum generalized NOON states provide maximal advantage in the task of two-way communication, just as the single-photon entangled state, by maximally violating the GYNI inequality (\ref{eq:gyni}). 
In line with this result, we have studied the robustness of the even-coherent state, when subjected to models of lossy apparatuses, and have shown it to outperform the single-photon entangled state in certain regimes of the parameters involved, although the single-photon entangled state, may in turn, outperform the even-coherent NOON state in some other range of values of the same parameters. On using the model of the lossy beam splitter, we see that the even-coherent NOON state outperforms (in the regime of low average photon number and low quantum efficiency $\eta$) the single-photon entangled state. On application of the the lossy detector, using the Tan-Krivitsky-Englert \cite{Tan16} model, the results in both states are comparable (maximal violation) when the efficiency is close to unity and saturation and the average photon number of the even-coherent NOON state are high. The fact that the even-coherent NOON state, in the regime of high detector efficiency $\kappa$, ends up exhibiting some violation (not maximal) even when its average photon number is low, irrespective of the saturation number, is noteworthy. Using the Sperling-Vogel-Agarwal \cite{Sperling2012} model, we observe that in the regime of low average photon number and low noise ($\nu$), for all values of saturation, the even-coherent NOON state outperforms the single-photon entangled as it exhibits a higher violation of GYNI when the detector efficiency $\kappa$ is around 0.5. As a detour, we have also studied the non-locality of some of these generalized NOON states using reference frame independent measurements. The results of our study prompt us to hypothesize that only the generalized NOON state in which $\ket{\xi}$ is non-classical will furnish a violation of the Bell-type inequality using reference frame independent measurements. In case such a hypothesis holds good, non-classicality in the single-mode state $\ket{\xi}$-- embedded in the two-mode state $\ket{\Phi}$ of eq. (\ref{eq:NOON})-- would play a pivotal role in acheiving two-way communication and in exhibiting non-locality in a reference-frame independent manner.

The theorem (c.f. section \ref{subsec:ncl}) is likely to remain a mathematical artefact unless one finds feasible ways of preparing any generalized NOON state. We have indicated towards possible ways of preparing even-coherent NOON states. The prospect of preparing other generalized NOON states is the subject of quantum state engineering (for instance, refer to \cite{Kovlakov18} and references therein). We have investigated the performance of even-coherent NOON state under noisy BS and noisy detectors separately. Taking into account the presence of noise at each step simultaneously is likely to put the existing models of loss to test and may provide a better comparison between generalized NOON state and the single photon two mode entangled state. In line with the results that we have presented, the imminent step is to study multiparty multi-way communication. Recent progress in single electron sources \cite{BFB+13,NOC+07,JCS+03}, and an experimental demonstration of single electron entanglement and non-locality \cite{DBB+16} motivate to ask of the possibility of extending the questions addressed in our work to two- or multi-way communication using entangled states of single massive particles. 
Moreover, our present work leads one to potentially address the possibility of quantifying the non-classicality in any single-mode pure state $\ket{\xi}$, embedded in a two-mode entangled state $\ket{\Phi}$ (eq. (\ref{eq:NOON})), via the maximum possible amount of violation of the GYNI inequality (\ref{eq:gyni}) that the generalized NOON state furnishes.

\section{Acknowledgements}
We, the authors, would like to express our gratitude towards Sibasish Ghosh for his valuable inputs and for having discussed at length with us. BG and SA acknowledge The Institute of Mathematical Sciences, Chennai for hosting them for a period during the tenure of which, this work was conceived.

\appendix

\section{Average photon number of the even-coherent NOON state}
\label{ap:avg}
The average photon number $\mathcal{\Tilde{N}}$ of the state $\ket{\Phi_E}$ (eq. (\ref{eq:evco})) obtained from
$\bra{\Phi_E}{\boldsymbol{\Hat{N}}}_A\otimes\boldsymbol{I}+\boldsymbol{I}\otimes\boldsymbol{\Hat{N}}_B\ket{\Phi_E}$ as a function of $\alpha$. 
\begin{equation}
\mathcal{\Tilde{N}} = \frac{|\alpha|^2}{(1+e^{-|\alpha|^2})^2},
\end{equation}
where ${\boldsymbol{\Hat{N}}}_A$ (${\boldsymbol{\Hat{N}}}_B$) is the number operator corresponding to mode $A$ ($B$). Notice that $\mathcal{\Tilde{N}}$ is merely a factor times the average photon number of the state $\ket{\alpha_e}$.
It is evident from the above equation that the average photon number of $\ket{\Phi_E}$ is less than 1 \textit{iff} $|\alpha|^2$ is less than 1.
%%%%%%%%%%%%%%%%%%%%%%%%%%%%%%%%%%%%%%%%%%%%%%%%%%%%%%%%%%%%%%%%%%%%%%%
\section{Generalized NOON states made out of squeezed vacuum states}
\label{ap:sq}
Let's begin with some preliminary notions about the squeezed vacuum states. The squeezed vacuum state $\ket{\xi}  = \hat{S}(\xi) \ket{0} $ is obtained 
by operating the squeezing operator $\hat{S}(\xi)$  
\begin{equation}
\hat{S}(\xi) = \exp [\frac{1}{2} (\xi^{\ast} a^2 - \xi a^{\dagger 2})]
\label{eq:sqz}
\end{equation}
on the vacuum state $\ket{0}$, where  $\xi = re^{i\theta}$, $r$ is the squeezing parameter such that $0\leq r < \infty$ and $0\leq \theta \leq 2 \pi$. The squeezed vacuum state, in terms of Fock states, is expressed as follows: 

\begin{equation}
\ket{\xi_{sq}} {\equiv} \hat{S}(\xi)\ket{0} = \frac{1}{\sqrt{\cosh{r}}} \sum_{m=0}^\infty (-1)^m \frac{\sqrt{(2m)!}}{2^m m!} e^{im\theta} (\tanh{r})^m \ket{2m}
\label{eq:sqzd}
\end{equation}

The important point to specify here is that the average photon number of squeezed vacuum  state is 
$\expval{\Hat{N}}_{\xi_{sq}} \equiv \expval{a^\dagger a}{\xi_{sq}} = (\sinh{r})^2 $. Thus, although we are using the squeezed state, the average photon number can, in principle, be made comparable to lowered below 1. 

Let us now apply the protocol given in section \ref{sec:proto} to the squeezed vacuum NOON state $ \ket{\Phi_{SQ}} = \frac{1}{\sqrt{N(\xi_{sq})}} (\ket{\xi_{sq}}\ket{0} + \ket{0}\ket{\xi_{sq}})$, where $N(\xi_{sq})$ is the normalization factor. The average photon number of the squeezed vacuum NOON state is a factor times that of the squeezed vacuum state. After encoding the information of the inputs, as explained in the main text, the state will be 

\begin{equation}
\ket{\Phi^{\prime}_{SQ{xy}}} = \frac{1}{\sqrt{N_{xy}(\xi_{sq})}} [(-1)^x\ket{\xi_{sq}}\ket{0} + (-1)^y\ket{0}\ket{\xi_{sq}}],
\label{eq:sqzent}
\end{equation}
where $N_{xy}(\xi_{sq})$ is the modified normalization factor (input dependent).
Alternately, it may be expressed in the number basis as 

	\begin{equation}
	\ket{\Phi^{\prime}_{SQ{xy}}} = \frac{1}{\sqrt{N_{xy}(\xi_{sq})\cosh{r}}}\sum_{m=0}^\infty \frac{(-\tanh{r})^m}{2^m m!}\sqrt{(2m)!} [(-1)^x \ket{2m,0} + (-1)^y \ket{0,2m}].
	\label{eq:sqzdnm}
	\end{equation}

After the action of the 50:50 BS, on the state (\ref{eq:sqzdnm}), for given inputs $x,y \in \{0,1\}$, the output states 
can be written as 

	\begin{align}
	\ket{\Phi_{SQ{00}}^{\prime{\prime}}} = -\ket{\Phi_{SQ{11}}^{\prime{\prime}}} = \frac{1}{\sqrt{2(\cosh{r}+1)}} \sum_{m=0}^\infty \sum_{k=0}^{m} \frac{(-\tanh{r})^m}{2^{2m-1} m!}\nonumber\\
	 {\times}\sqrt{(2m)!} {\left(\begin{array}{c} 2m \\ 2k \end{array}\right)}^{\frac{1}{2}} \ket{2k, 2(m-k)}, \\ \ket{\Phi_{SQ{01}}^{\prime{\prime}}} = -\ket{\Phi_{SQ{10}}^{\prime{\prime}}} =  \frac{1}{\sqrt{2(\cosh{r}+1)}} \sum_{m=0}^\infty \sum_{k=0}^{m} \frac{(-\tanh{r})^m}{2^{2m-1} m!}\nonumber\\
	  {\times}\sqrt{(2m)!} {\left(\begin{array}{c} 2m \\ 2k - 1 \end{array}\right)}^{\frac{1}{2}} \ket{2k -1, 2(m-k)+1}.
	\label{eq:sqbig}
	\end{align}

By using the measurements given in eq. (\ref{eq:meas}) of the main text and from eq. (\ref{eq:sqbig}) the associated correlations take values as $P(00|00) = P(11|01) = P(11|10) = P(00|11) = 1$ and thereby violate the GYNI inequality maximally. 
%%%%%%%%%%%%%%%%%%%%%%%%%%%%%%%%%%%%%%%%%%%%%%%%%%%%%%%%%%%%%%%%%%%
\section{Generalization}
\label{sec:gen}
In this section we prove that any generalized NOON state furnishes a violation of the GYNI inequality if and only if the single-mode pure state embedded in it is non-classical and subsequently provide some examples. 
%%%%%%%%%%%%%%%%%%%%%%%%%%%%%%%%%%%%%%%%%%%%%%%%%%%%%%
\subsection{Proof of Theorem \ref{thm1}}
\label{subsec:proof}
We call $\ket{\Phi_{xy}}$, the phase-encoded version of the generalized NOON state $\ket{\Phi}$, in which the embedded single-mode pure state is $\ket{\xi}$ (refer to eq. (\ref{eq:NOON})).
The phase encoded state, on passing through a 50:50 beam splitter, gives the following output states on the basis of parity of the inputs $x$ and $y$.
\begin{align}
	\ket{\Phi_{00}} = -\ket{\Phi_{11}}
	&=\frac{2}{\sqrt{2(1+|\lambda_0|^2)}}\sum^{\infty}_{n=0}\sum^{\infty}_{k=0}\frac{\lambda_{n+2k}}{2^{\frac{n+2k}{2}}}{\binom{n+2k}{2k}}^{\frac{1}{2}}\ket{n}\ket{2k} \label{gen_cor},\\
	\ket{\Phi_{01}} = -\ket{\Phi_{10}}
	&=\frac{2}{\sqrt{2(1-|\lambda_0|^2)}}\sum^{\infty}_{n=0}\sum^{\infty}_{k=0}\frac{\lambda_{n+2k+1}}{2^{\frac{n+2k+1}{2}}}{\binom{n+2k+1}{2k+1}}^{\frac{1}{2}}\ket{n}\ket{2k+1}. \label{gen_anticor}
\end{align}	
It is evident from equations (\ref{gen_cor}) and (\ref{gen_anticor}), that the Bob's choice of measurement must be 
\begin{equation}
\mathcal{M^{(B)}}=\{\Pi^B_0=\sum^{\infty}_{m=0}\dyad{2m}{2m},\Pi^B_1=\sum^{\infty}_{m=0}\dyad{2m+1}{2m+1}\},
\end{equation}
for the purpose of extracting information about Alice's input (required for the violation of the GYNI inequality).
We are now required to find a suitable choice of measurement $\mathcal{M^{(A)}}=\{\Pi^A_0,\Pi^A_1\}$, for Alice. Recall from the cases that have been discussed so far, the LHS of the GYNI inequality that we've been studying, essentially is a sum of two terms. 
\begin{equation}
\bra{\Phi_{00}}(\Pi^A_0{\otimes}\Pi^B_0)\ket{\Phi_{00}}+\bra{\Phi_{01}}(\Pi^A_1{\otimes}\Pi^B_1)\ket{\Phi_{01}}
{\equiv}F(\xi)
\label{eq:big_eq}
\end{equation} 
For the sake of simplifying our calculations, we define the normalization constant after phase encoding, as $\mathcal{N}_{xy}=2(1+(-1)^{x+y}|\lambda_0|^2)$.
By making use of equations (\ref{gen_cor}) and (\ref{gen_anticor}) in equation (\ref{eq:big_eq}), on performing some algebraic manipulations, we find
\begin{align}
	F(\xi)
	&=\frac{4}{\mathcal{N}_{00}}\sum^{\infty}_{n=0}\sum^{\infty}_{k=0}\frac{|\lambda_{n+2k}|^2}{2^{n+2k}}\binom{n+2k}{2k}\nonumber\\
	&+4\sum^{\infty}_{k=0}\frac{\mathcal{M}_{2k+1}}{\mathcal{N}_{01}}\bra{\eta_{2k+1}}\Pi^A_1\ket{\eta_{2k+1}}\nonumber\\
	&-4\sum^{\infty}_{k=0}\frac{\mathcal{M}_{2k}}{\mathcal{N}_{00}}\bra{\eta_{2k}}\Pi^A_1\ket{\eta_{2k}}.
	\label{eq:big_mod}
\end{align}
Here $\ket{\eta_k}=\frac{1}{\sqrt{\mathcal{M}_k}}\sum^{\infty}_{n=0}\frac{\lambda_{n+k}}{2^{\frac{n+k}{2}}}{\binom{n+k}{k}}^{\frac{1}{2}}\ket{n}$ and $\mathcal{M}_k = \sum^{\infty}_{n=0}\frac{|\lambda_{n+k}|^2}{2^{n+k}}\binom{n+k}{k}$. The first term in equation (\ref{eq:big_mod}) can be found to be $1$ after a few steps of algebra. In order to make sure that the negative term in equation (\ref{eq:big_mod}) doesn't contribute, we must choose $\Pi^A_1$ to be orthogonal to all the states $\ket{\eta_{2k}}$ for $k=0,1,2,..$. It can then be argued that $F(\xi)$ is always greater than $1$. 

Let us take a closer look at the sum of the second and the third terms of equation (\ref{eq:big_mod}).
\begin{align}
	4\sum^{\infty}_{k=0}\frac{\mathcal{M}_{2k+1}}{\mathcal{N}_{01}}\bra{\eta_{2k+1}}\Pi^A_1\ket{\eta_{2k+1}}
	-4\sum^{\infty}_{k=0}\frac{\mathcal{M}_{2k}}{\mathcal{N}_{00}}\bra{\eta_{2k}}\Pi^A_1\ket{\eta_{2k}}\nonumber\\
	=4[\sum^{\infty}_{k=0}\frac{1}{\mathcal{N}_{01}}\bra{\tilde{\eta}_{2k+1}}\Pi^A_1\ket{\tilde{\eta}_{2k+1}}
	-\sum^{\infty}_{k=0}\frac{1}{\mathcal{N}_{00}}\bra{\tilde{\eta}_{2k}}\Pi^A_1\ket{\tilde{\eta}_{2k}}] \nonumber\\
	=4\mathcal{N}^{\prime}\sum^{\infty}_{k=0}[\frac{1}{\mathcal{N}_{01}}\frac{\hat{a}^{2k+1}}{\sqrt{(2k+1)!}}\dyad{\phi}{\phi}\frac{\hat{a}^{{\dagger}2k+1}}{\sqrt{(2k+1)!}}\nonumber\\
	-\frac{1}{\mathcal{N}_{01}}\frac{\hat{a}^{2k}}{\sqrt{(2k)!}}\dyad{\phi}{\phi}\frac{\hat{a}^{{\dagger}2k}}{\sqrt{(2k)!}}],
\end{align}
where $\ket{\tilde{\eta}_k}=\sum^{\infty}_{n=0}\frac{\lambda_{n+k}}{2^{\frac{n+k}{2}}}{\binom{n+k}{k}}^{\frac{1}{2}}\ket{n}$, $\mathcal{N^{\prime}}=\sum^{\infty}_{n=0}\frac{|\lambda_n|^2}{2^n}$, and $\ket{\phi}=\frac{1}{\sqrt{\mathcal{N^{\prime}}}}\sum^{\infty}_{n=0}\frac{\lambda_n}{2^{\frac{n}{2}}}\ket{n}$.

All the (un-normalized) states $\hat{a}^k\ket{\phi}$ for $k=0,1,2,...$ are linearly independent as long as $\ket{\phi}$ is not the coherent state \footnote{If $\ket{\phi}$ is a linear superposition of finitely many Fock states, which happens only when $\ket{\xi}$ is a linear superposition of finitely many Fock states, these un-normalized states vanish for a large enough $k$. The ones that don't vanish are linearly independent. When $\ket{\xi}$ is  a coherent state (say $\ket{\lambda}$), then $\hat{a}^k\ket{\phi}$s become proportional to one and the same state $\ket{\frac{\lambda}{\sqrt{2}}}$ for all $k=0,1,2,..$}. Thus, it should be possible, in principle, to choose at least one $\ket{\eta}$ which is orthogonal to $\hat{a}^{2k}\ket{\phi}$ for all $k=0,1,2,..$. As $\hat{a}^{2k+1}\ket{\phi}$s are linearly independent of $\hat{a}^{2k}\ket{\phi}$ (for $k=0,1,..$), $\ket{\eta}$ will have a non-zero overlap with the subspace spanned by all the $\hat{a}^{2k+1}\ket{\phi}$s.
We then choose $\Pi^A_1=\dyad{\eta}{\eta}$.
Using the afore-mentioned argument, we may show that 
\begin{equation}
F(\xi) = 1+ \frac{4}{\mathcal{N}_{01}}\sum^{\infty}_{k=0}|\braket{\eta}{\tilde{\eta}_{2k+1}}|^2{\geq}1.
\label{eq:f_final}
\end{equation}
It can be deduced from equation (\ref{eq:big_mod}) that if $\ket{\psi}$ is a coherent state, irrespective of the choice of $\Pi^A_1$, $F(\xi)=1$ and hence there is no violation of the GYNI inequality under consideration. As mentioned previously, the second term in eq. (\ref{eq:f_final}) is positive whenever $\ket{\xi}$ is not a coherent state. It is worth mentioning here that the second term in eq. (\ref{eq:f_final}) can be tuned by choosing $\Pi^A_1$ accordingly\footnote{$\Pi^A_1$ can be chosen to be a higher rank projector-- $\sum_l\dyad{{\eta^l}}{\eta^l}$, with $\braket{\eta^l}{\eta^{l{\prime}}}=\delta_{ll^{\prime}}$ and each $\ket{\eta^l}$ being orthogonal to all the states $\hat{a}^{2k}\ket{\phi}$.}. This in turn, would enable maximum possible violation for the given state $\ket{\Phi}$.  
\begin{corollary}
	All finite-dimensional states (FDSs) beget generalized NOON states which violate at least the GYNI inequality (\ref{eq:gyni}).
\end{corollary}
\textbf{Proof:}
The proof is similar to the one that has just been described and makes use of the argument in \cite{Note3}.
%%%%%%%%%%%%%%%%%%%%%%%%%%%%%%%%%%%%%%%%%%%%%%%%%%%%%
\subsection{An example in which $L$ is infinite \label{ap:subsec_inf}}
Existing literature shows that the photon added coherent state is non-classical (see, for example, \cite{Park2017}). This follows from the fact that all finite-dimensional states (FDSs) are non-classical and the criteria of the demarginalization maps (DMs) posed in \cite{Park2017}, are invariant under displacement operations. 
The generalized NOON state made out of the photon added coherent state is of the following form.
\begin{equation}
\ket{\Phi^3}_{AB} = \frac{1}{\sqrt{\mathcal{N}^{\prime}}} (\hat{a}^{\dagger}\ket{\alpha}_A \ket{0}_B + \ket{0}_A \hat{b}^{\dagger}\ket{\alpha}_B),
\label{eq:pan}
\end{equation}
where $\mathcal{N}^{\prime}$ is the normalization factor, $\hat{a}^{\dagger}$, and $\hat{b}^{\dagger}$ are the creation operators for modes $A$ and $B$ respectively. This state on passing through a 50:50 beam splitter (post phase encoding) can be grouped into two categories, namely correlated and anti-correlated.
\paragraph{When the inputs are correlated}
\begin{align}
	\ket{\Phi^{3\prime}_{00}}&=-\ket{\Phi^{3\prime}_{11}} \nonumber\\
	&= \frac{e^{\frac{-|\alpha|^2}{4}}}{\sqrt{2(1+|\alpha|^2)}}\sum^{\infty}_{m=0}\frac{\alpha^{2m-1}}{2^{m-1}\sqrt{(2m)!}}(2m\mathbf{I}+\frac{\alpha}{\sqrt{2}}\hat{a}^{\dagger})\ket{\frac{\alpha}{\sqrt{2}}}\ket{2m}
	\label{eq:pa_corr2}
\end{align}
It is important to note that the state in mode B has its entire support lying inside the subspace spanned by the even Fock states. Let us call the resultant density matrix of mode A as $\rho^{3\prime}_{corr}$. We shall shortly return to discussing the significance of this expression. Let us first study the expression for the output state when the inputs are anti-correlated.	
\paragraph{When the inputs are anti-correlated}
\begin{align}
	\ket{\Phi^{3\prime}_{01}}&=-\ket{\Phi^{3\prime}_{10}} \nonumber\\
	&= \frac{e^{\frac{-|\alpha|^2}{4}}}{\sqrt{2(1+|\alpha|^2)}}\sum^{\infty}_{m=0}\frac{\alpha^{2m}}{2^{m-\frac{1}{2}}\sqrt{(2m+1)!}}((2m+1)\mathbf{I}+\frac{\alpha}{\sqrt{2}}\hat{a}^{\dagger})\ket{\frac{\alpha}{\sqrt{2}}}\ket{2m+1}
	\label{eq:pa_acorr2}
\end{align}
Once again, it is significant to note that the state in mode B has its support lying entirely inside the subspace spanned by odd Fock states. Let us call the resultant density matrix of mode A as $\rho^{3\prime}_{anti-corr}$. We shall demonstrate that although $\rho^{3\prime}_{corr}$ and $\rho^{3\prime}_{anti-corr}$ appear to have the same supports, it is possible to choose a measurement on Alice's subsystem which allows her to distinguish her states in the two aforementioned cases. Such an argument holds good because Alice's Hilbert space is infinite dimensional.
\paragraph{Furnishes violation of a GYNI inequality:} 
It is evident from equations (\ref{eq:pa_corr2}) and (\ref{eq:pa_acorr2}) that Bob must choose the following measurement on his subsystem.
\begin{equation}
\mathcal{M^{(B)}} {\equiv} \{\Pi_0^{(B)} = \sum^{\infty}_{m=0}\dyad{2m}{2m},\Pi_1^{(B)} = \sum^{\infty}_{m=0}\dyad{2m+1}{2m+1}\}
\end{equation}
Let us note that normalized version of the state $a^{\dagger}\ket{\frac{\alpha}{\sqrt{2}}}$ can be written as a superposition of the following two states
\begin{equation}
\frac{a^{\dagger}\ket{\frac{\alpha}{\sqrt{2}}}}{\sqrt{1+\frac{|\alpha|^2}{2}}} = \frac{\alpha^*}{\sqrt{2+|\alpha|^2}}\ket{\frac{\alpha}{\sqrt{2}}} + \sqrt{\frac{2}{2+|\alpha|^2}}\ket{\psi^{\perp}(\alpha)},
\end{equation}
where $\ket{\psi^{\perp}(\alpha)}$ is normalized and is orthogonal to $\ket{\frac{\alpha}{\sqrt{2}}}$. 
Now on Alice's subsystem we choose the projective measurement $\mathcal{M^{(B)}}=\{\Pi^{(A)}_0,\Pi^{(A)}_1\}$ such that $\ket{\frac{\alpha}{\sqrt{2}}}$ lies entirely in the support of $\Pi^{(A)}_1$, while $\ket{\psi^{\perp}(\alpha)}$ lies entirely in the support of the orthogonal projector $\Pi^{(A)}_0$. Barring these constraints, Alice has complete freedom to choose her projectors. On the basis of these constraints alone, the following can be established after some algebraic manipulations and on assuming that $\alpha{\neq}0$.
\begin{equation}
\bra{\Phi^{3\prime}_{00}}(\Pi_0^{(A)}{\otimes}\Pi_0^{(B)})\ket{\Phi^{3\prime}_{00}} = \frac{1}{2(1+|\alpha|^2)}(1+e^{-|\alpha|^2})
\label{eq:co}
\end{equation}
\begin{equation}
\bra{\Phi^{3\prime}_{01}}(\Pi_1^{(A)}{\otimes}\Pi_1^{(B)})\ket{\Phi^{3\prime}_{01}} =
\frac{1}{2(1+|\alpha|^2)}(1+2|\alpha|^2+e^{-|\alpha|^2})
\label{eq:aco}
\end{equation}
On choosing $\mathcal{M^{(A)}},\mathcal{M^{(B)}}$, the the GYNI inequality quite simply is
\begin{equation}
\bra{\Phi^{3\prime}_{00}}(\Pi_0^{(A)}{\otimes}\Pi_0^{(B)})\ket{\Phi^{3\prime}_{00}}+\bra{\Phi^{3\prime}_{01}}(\Pi_1^{(A)}{\otimes}\Pi_1^{(B)})\ket{\Phi^{3\prime}_{01}}{\leq}1
\label{ineq:mod_gyni}
\end{equation}
Using equations (\ref{eq:co}) and (\ref{eq:aco}), we find that the LHS of inequality (\ref{ineq:mod_gyni}) is $1+\frac{e^{-|\alpha|^2}}{1+|\alpha|^2}$ ($\geq 1$ for all finite values of $|\alpha|^2$ with equality holding only when $\alpha=0$). Hence, the photon added coherent NOON state always furnishes a violation of a GYNI inequality. Incidentally when $\alpha=0$, the photon added coherent NOON state (c.f eq. (\ref{eq:pan})) reduces to the single-photon entangled state and that this state furnishes a maximal violation of a GYNI inequality is well established. Note that the photon-added coherent NOON state violates the GYNI inequality maximally only when $\alpha=0$.
%%%%%%%%%%%%%%%%%%%%%%%%%%%%%%%%%%%%%%%%%%%%%%%%%%%%%%%%%%%%%
\subsection{An example in which $L$ is finite}
\label{ap:subsec_fin}
Let us consider the following non-classical state $\ket{\psi_1}$ which a superposition of $\ket{0}$ and $\ket{1}$.
\begin{equation}
\ket{\psi_1}=\sqrt{\lambda}\ket{0}+e^{\iota{\phi}}\sqrt{1-\lambda}\ket{1}
\label{new1}
\end{equation}
Here $0\leq{\lambda}<1$ and $\phi$ is a phase. When this condition on $\lambda$ is satisfied, the state $\ket{\psi_1}$ is always non-classical as it is a FDS.
The corresponding generalized NOON state, post phase encoding is given by:
\begin{equation}
\ket{\Phi^1_{xy}}=\frac{1}{\sqrt{N_{xy}(\lambda)}}[(-1)^x\ket{\psi_1{0}}+(-1)^y\ket{0{\psi_1}}],
\end{equation}
where, $N_{xy}(\lambda)=2[1+(-1)^{x+y}{\lambda}]$. On passing through a lossless 50:50 beam splitter this state undergoes a transformation and the resultant state is the following.
\begin{equation}
\resizebox{0.8\hsize}{!}{$\ket{\Phi^{1\prime}_{xy}}=\frac{1}{\sqrt{N_{xy}}}[\{(-1)^x+(-1)^y\}\{\sqrt{\lambda}\ket{00}_{AB}+e^{\iota{\phi}}\sqrt{\frac{1-\lambda}{2}}\ket{10}_{AB}\}+\{(-1)^x-(-1)^y\}e^{\iota{\phi}}\sqrt{\frac{1-\lambda}{2}}\ket{01}_{AB}]$}
\label{eq:postM}
\end{equation}
Now, in line with the measurements performed on the single photon state, we find the probability of Alice having a photon and Bob having none, when their inputs are correlated along with the probability of Alice having no photons and Bob having a photon, while their inputs are anti-correlated. Let us now formally define the measurements and the labels that we associate with the outcomes of the measurements, followed by the relevant correlations. Let us now define the outcomes according to the projector that click.
\begin{equation}
	\mathcal{M}^{\prime(A)} = \{a=1 {\equiv} \ket{0}\!\!\!_{_A}\!{\bra{0}}, a=0 {\equiv} \ket{1}\!\!\!_{_A}\!{\bra{1}}\}; \quad
	\mathcal{M}^{\prime(B)} = \{b=0 {\equiv} \ket{0}\!\!\!_{_B}\!{\bra{0}}, b=1 {\equiv} \ket{1}\!\!\!_{_B}\!{\bra{1}}\}
\end{equation} 
 We now write the expressions for the correlations of our concern
\begin{align}
	P(0,0|0,0)&=P(0,0|1,1)=\bra{\Phi^{\prime}_{00}}\dyad{1}{1}{\otimes}\dyad{0}{0}\ket{\Phi^{\prime}_{00}}=\frac{1-{\lambda}}{1+\lambda}\\
	P(1,1|0,1)&=P(1,1|1,0)=\bra{\Phi^{\prime}_{01}}\dyad{0}{0}{\otimes}\dyad{1}{1}\ket{\Phi^{\prime}_{01}}=1
\end{align}
Thus, the value (a particular value, say $\mathcal{J}_1$, of $\mathcal{J}$) of the inequality (\ref{eq:gyni}) has the following functional dependence on $\lambda$.
\begin{equation}
{\mathcal{J}}_1=\frac{1}{1+\lambda}>\frac{1}{2}{\hspace{1mm}} \text{as}\hspace{1mm} \lambda<1.
\label{J1}
\end{equation}
Thus, we observe that for the chosen set of measurements, inequality (\ref{eq:gyni}) is always violated. Here, ${\mathcal{J}}_1=1$ (maximum violation) when $\lambda=0$, as $\ket{\Phi^1_{xy}}$ then becomes the phase encoded version of the single-photon entangled state.

\begin{figure}[ht]
	\centering
	\includegraphics[width=10cm,height=8cm]{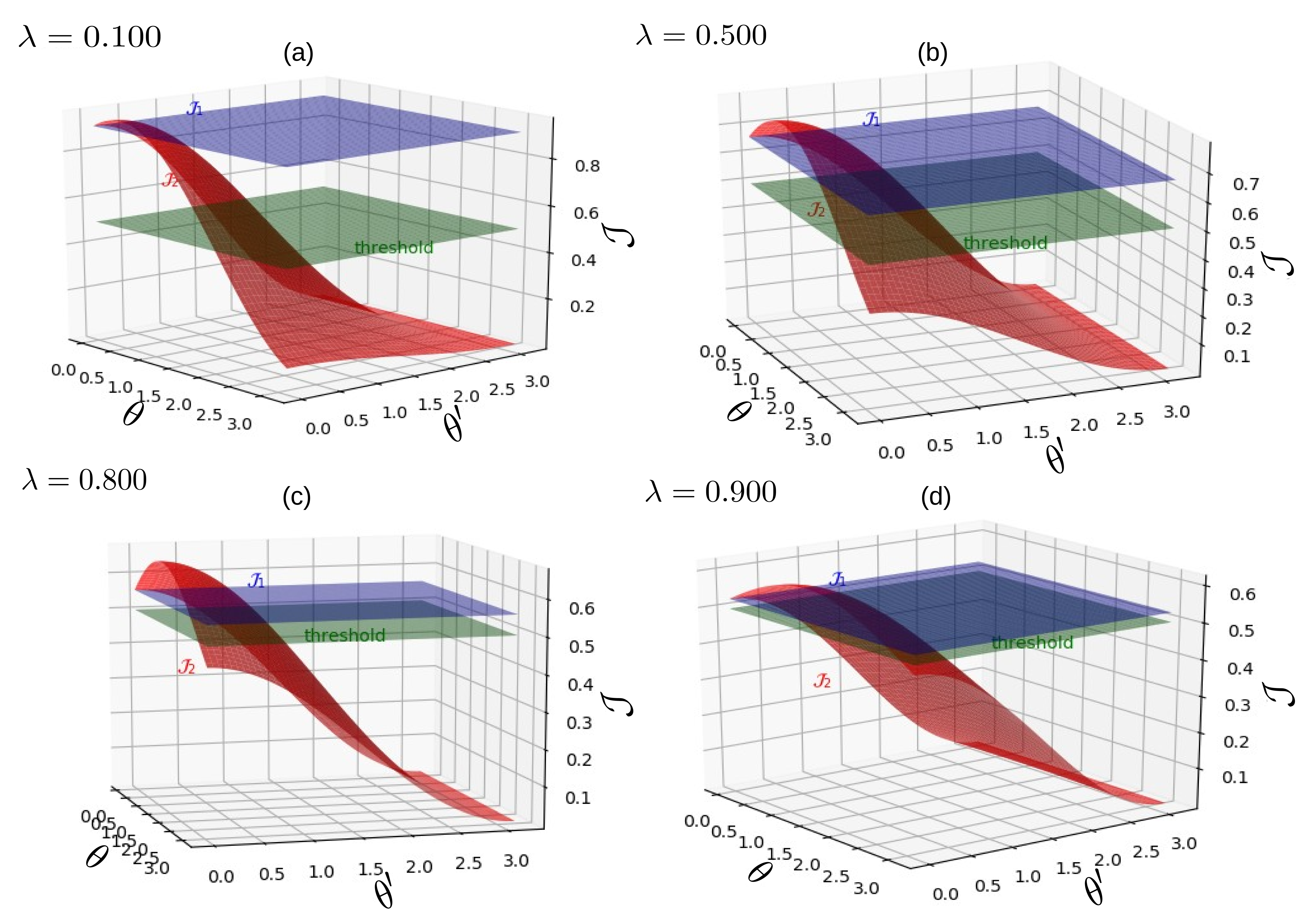}
	\caption{(Colour online) The green plane in each figure represents the threshold, $z=\frac{1}{2}$. The blue planes are the $z=\frac{1}{1+\lambda}$ planes ($\mathcal{J}_1$). The red surfaces are $\mathcal{J}_2$. It is intuitive that when $\lambda{\approx}1$, i.e., when the state becomes classical, neither of the measurements furnish any violation. This is reflected by the fact that when, $\lambda=0.9$, the green and the red surfaces barely cross the threshold. As the value of $\lambda$ drops from 1 and tends towards 0, the set $\{\mathcal{M}^{\prime}_A,\mathcal{M}^{\prime}_B\}$ allows us to observe a better violation that the set $\{\mathcal{M}^{\prime{\prime}}_A,\mathcal{M}^{\prime{\prime}}_B\}$, when $\theta,\theta^{\prime}\neq0$ for the latter set.}.
	\label{fig:ch31}
\end{figure}

We would like to find out if any other set of measurements can furnish a better violation than the aforementioned set, since a higher violation would increase our chances of winning the game. In order to achieve this in the most general way, we define two sets of mutually orthogonal states.
\begin{align}
	\ket{\chi}_A =\cos\frac{\theta}{2}\ket{0}_A+e^{\iota{\epsilon}}\sin\frac{\theta}{2}\ket{1}_A, \quad \ket{\chi^{\perp}}_A = \sin\frac{\theta}{2}\ket{0}_A-e^{\iota{\epsilon}}\cos\frac{\theta}{2}\ket{1}_A,\\
	\ket{\eta}_B = \cos\frac{\theta^{\prime}}{2}\ket{0}_A+e^{\iota{\epsilon^{\prime}}}\sin\frac{\theta^{\prime}}{2}\ket{1}_A,\quad \ket{\eta^{\perp}}_B = \sin\frac{\theta^{\prime}}{2}\ket{0}_A-e^{\iota{\epsilon^{\prime}}}\cos\frac{\theta^{\prime}}{2}\ket{1}_A,
\end{align}
where $\theta,\theta^{\prime}{\in}[0,\pi]$ and $\epsilon,\epsilon^{\prime}{\in}[0,2\pi]$ (phases).
We then define the following sets of measurements and the labels of the corresponding outcomes.
\begin{equation}
	\mathcal{M}^{\prime{\prime}{A}} = \{a=1 {\equiv} \ket{\chi}\!\!\!_{_A}\!{\bra{\chi}}, a=0 {\equiv} \ket{\chi^{\perp}}\!\!\!_{_A}\!{\bra{\chi^{\perp}}}\} \quad
	\mathcal{M}^{\prime{\prime}{B}} = \{b=0 {\equiv} \ket{\eta}\!\!\!_{_B}\!{\bra{\eta}}, b=1 {\equiv} \ket{\eta^{\perp}}\!\!\!_{_B}\!{\bra{\eta^{\perp}}}\}
\end{equation}
Note that the set $\{\mathcal{M}^{\prime{\prime}}_A,\mathcal{M}^{\prime{\prime}}_B\}$ reduces to the set $\{\mathcal{M}^{\prime}_A,\mathcal{M}^{\prime}_B\}$ when, $\theta,\theta^{\prime}=0$.
On calculating the LHS of inequality (\ref{eq:gyni}) of the , by performing these measurements on the state $\ket{\Phi^{\prime}_{xy}}$ (eq. \ref{eq:postM}), we obtain
\begin{equation}
{\mathcal{J}}_2 = \frac{\cos^2\frac{\theta^{\prime}}{2}}{1+\lambda}\left[\lambda\sin^2\frac{\theta}{2}+\cos^2\frac{\theta}{2}+\sin{\theta}\sqrt{\frac{\lambda(1-\lambda)}{2}}\right]
\label{J2}
\end{equation}
We have compared the surface $\mathcal{J}_2$ with the plane $\mathcal{J}_1$, for all values of $\theta$ and $\theta^{\prime}$, for given values of $\lambda$. Fig. (\ref{fig:ch31}) summarizes our results.
Barring the cases when $\lambda=0$ and $\lambda=1$, the second set of measurements allow for a higher violation of the chosen inequality for some values of $\theta$ and $\theta^{\prime}$. The maximum value of the GYNI inequality furnished by the state under the second set of measurements, for a particular value of $\lambda$ is found by numerically scanning over values of $\theta$ and $\theta^{\prime}$. When $\lambda=0.5$, the maximum value is found to be $\approx 0.71$.
%%%%%%%%%%%%%%%%%%%%%%%%%%%%%%%%%%%%%%%%
\section{More on the lossy beam splitter (BS)}
\label{ap:lossy_bs}
In this section, we shall describe the action of the lossy beam splitter (discussed in section \ref{subsec:bs_loss}) on a four-mode input state. We shall follow this up with a description of an alternate model of the lossy BS.
%%%%%%%%%%%%%%%%%%%%%%%%%%%%%%%%%%%%%%%%%%%%%%%%%%%
\subsection{Action of the lossy beam splitter on the four-mode Fock states }
\label{ap:subsec_bs_extra}
Let us now illustrate the formalism discussed in section \ref{subsec:bs_loss}, by considering transformation of the four mode Fock state under the action of the lossy beam splitter. An interested reader is requested to refer to section IV of \cite{KND98} for further details. Let
\begin{equation}
\Hat{\rho}_{in} = \ketbra{\psi_{in}}{\psi_{in}},
\end{equation}
with
\begin{equation}
\ket{\psi_{in}} = \ket{n_1,n_2,n_3,n_4} = \prod_{\nu=1}^{4}\frac{\Hat{\alpha}^{\dagger{n_{\nu}}}_{\nu}}{\sqrt{n_{\nu}!}}\ket{0},
\end{equation}
be the density operator of the system in the case when $n_1$ and $n_2$ quanta are the field mode excitations and $n_3$ and $n_4$ quanta are device mode excitations. On applying eq. (\ref{eq:u4}) on $\Hat{\rho}_{in}$, we obtain
\begin{equation}
\Hat{\rho}_{out} = \ketbra{\psi_{out}}{\psi_{out}},\quad 
\ket{\psi_{out}} = \prod_{\nu=1}^{4}\frac{\left(\sum_{\mu=1}^4{\Lambda}_{\mu{\nu}}\Hat{a}^{\dagger}_{\mu}\right)^{n_{\nu}}}{\sqrt{n_{\nu}!}}\ket{0},
\end{equation}
where $\Lambda_{\mu{\nu}}$ are the elements of the $4{\times}4$ unitary matrix of the lossy beam splitter. The term in brackets raised to the exponent $n_{\nu}$ can be expanded using multinomial expansion. The density operator of the outgoing field modes is then obtained by tracing out over the device modes.
\begin{equation}
\Hat{\rho}^{(F)}_{out} = Tr^{(D)}\{\ketbra{\psi_{out}}{\psi_{out}}\}.
\end{equation}
Based on our choices of $\mathbf{T}$ and $\mathbf{A}$ (eq. (\ref{eq:absor})), the following is the form of $\mathbf{\Lambda}$ used in our calculations.
\begin{equation}
\boldsymbol{{\Lambda}} = \frac{1}{\sqrt{2}}\begin{pmatrix}
\sqrt{\eta} & \sqrt{\eta} & \sqrt{1-\eta} & \sqrt{1-\eta}\\
\sqrt{\eta} & -\sqrt{\eta} & -\sqrt{1-\eta} & \sqrt{1-\eta}\\
-\sqrt{1-\eta} & -\sqrt{1-\eta} & \sqrt{\eta} & \sqrt{\eta}\\
-\sqrt{1-\eta} & \sqrt{1-\eta} & -\sqrt{\eta} & \sqrt{\eta}
\end{pmatrix}
\label{ap:lambda}
\end{equation}
%%%%%%%%%%%%%%%%%%%%%%%%%%%%%%%%%%%%%%%%%%%%%%%%%%%%%%%%%%
\subsection{Alternate description of the lossy beam splitter}
\label{ap:subsec_alt}
\begin{figure}[ht]
	\centering
	\includegraphics[width=8cm,height=8cm]{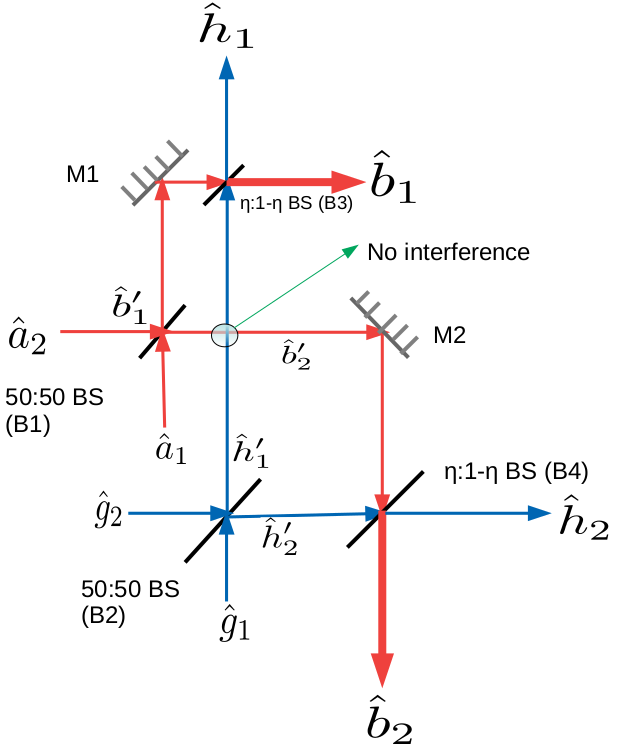}
	\caption{(Colour Online) The model proposed in \cite{KND98} (used here), in conjunction with the model of loss or gain proposed in \cite{Tischler2018} can be used to further understand the mixing of the field and the device modes in order to produce the net effect of loss. M1 and M2 are mirrors. B1 and B2 are 50:50 beam splitters. B3 and B4 are $\eta:1-\eta$ beam splitters up to overall phases. The outputs of B1 and B2 are made to interact with each other by passing through B3 and B4. The final output modes of the field, marked in bold, are of interest to us.} 
	\label{fig:new2}
\end{figure}

Off late, a development \cite{Tischler2018} proposes methods to realize arbitrary linear transformations allowing for both loss and gain. A lossy beam splitter, though no longer a unitary operation, is certainly a linear transformation. The input and output modes can be augmented with ancilla modes (device modes) and singular value decomposition of the full network can be performed which allows each component to be further decomposed into a series of elementary operations. Fig. (\ref{fig:new2}) represents the results of the following steps pictorially.

\begin{equation}
\begin{pmatrix}
\hat{a}_1 \\           
\hat{a}_2\\
\hat{g}_1 \\
\hat{g}_2 
\end{pmatrix}
\rightarrow
\begin{pmatrix}
\hat{b}^{\prime}_1 = \frac{\hat{a}_1+\hat{a}_2}{\sqrt{2}} \\           
\hat{b}^{\prime}_2 = \frac{\hat{a}_1-\hat{a}_2}{\sqrt{2}} \\
\hat{h}^{\prime}_1 = \frac{\hat{g}_1+\hat{g}_2}{\sqrt{2}} \\
\hat{h}^{\prime}_2 = \frac{\hat{g}_1-\hat{g}_2}{\sqrt{2}}
\end{pmatrix}
\rightarrow
\begin{pmatrix}
\hat{b}_1 = \sqrt{\eta}{\hat{b}^{\prime}_1} + \sqrt{1-\eta}\hat{h}^{\prime}_1\\           
\hat{b}_2 = \sqrt{\eta}{\hat{b}^{\prime}_2} - \sqrt{1-\eta}\hat{h}^{\prime}_2 \\
\hat{h}_1 = -\sqrt{1-\eta}{\hat{b}^{\prime}_1} + \sqrt{\eta}{\hat{h}^{\prime}_1} \\
\hat{h}_2 = -\sqrt{1-\eta}{\hat{b}^{\prime}_1} - \sqrt{\eta}{\hat{h}^{\prime}_1} 
\end{pmatrix}
\end{equation}
It can be verified that the four-mode output is identical to that which results from the action of eq. (\ref{ap:lambda}) on the four-mode input. This alternate description can also be used to explain the mixing between the input modes of the field and those of the device and results precisely in the model described in the section {\ref{subsec:bs_loss}}.
%%%%%%%%%%%%%%%%%%%%%%%%%%%%%%%%%%%%%%%%%%%%%%%%%%%%%%%%%%%%%%%%%%%%
\section{Tan-Krivitsky-Englert model of lossy PNRDs}
\label{ap:tan}
A model for photon detection with loss and saturation has been given in ref. \cite{Tan16}. Suppose a detector can resolve upto $N$ photons, this detector can then be modelled by a positive operator-valued measure (POVM) with $N+1$ outcomes: $\{\Pi_0,\Pi_1,...,\Pi_N\}$, satisfying completeness, $\sum_{i=0}^{N}\Pi_i = I$, and positivity, $\Pi_i\geq0$, for all $i=0,...,N$. For a perfect detector with no loss (unit quantum efficiency $\kappa$) and no saturation, $\{\Pi_i = \ketbra{i}{i}, i=1,2,...\}$. Let us define the $m$-th POVM effect of a lossy detector with quantum efficiency $\kappa$ is 
\begin{equation}
\Pi_{m} = \sum_{n=m}^{\infty}w_{m,n}(\kappa)\ketbra{n}{n},
\label{eq:povm}
\end{equation}
with 
\begin{equation}
w_{m,n}(\kappa) = {\kappa}^m(1-\kappa)^{n-m}\binom{n}{m},
\label{eq:povm2}
\end{equation}
for $m=0,1,2...$. Note in eq. (\ref{eq:povm}) that when $m$ photons arrive at the detector, there exists a non-zero probability, if $\kappa$ is not unity, for the detector to count more than $m$ photons. Thus, $\Pi_0 = \sum_{n=0}^{\infty}(1-\eta)^{n}\ketbra{n}{n}$. This implies that even in the absence of incoming photons, an inefficient detector might click. This precisely is what the term 'dark count' refers to. The effect of saturation prevents the detector from resolving between $N$ and $N+1$ (or more) photons. Thus we have for the $N$-th outcome:
\begin{equation}
\Pi_N = I-\sum_{m=0}^{N-1}\Pi_m.
\end{equation}
When $N=1$, this POVM reduces to $\{\Pi_0,\Pi_1\}$, where $\Pi_1=I-\Pi_0$.

Recall that for the even-coherent NOON state, the measurements defined in eq. (\ref{eq:meas}), have been course grained from all possible outcomes of lossless, photon number resolving detectors. We use a similar approach to group the aforementioned effects into two broad effects, which in the limit of $\kappa$ going to $1$ and $N$ going to infinity, would reduce to eq. (\ref{eq:meas}). We do this in two case: when the saturation is even and when the saturation is odd.
\subsubsection*{Case 1: When the detector can resolve upto $2N$ photons}
We define the following POVM for our state:
\begin{equation}
\overline{\Pi}^{(2N)}_{e} = \sum_{m=0}^{N-1}\Pi_{2m} + \frac{1}{2}(I-\sum_{m=0}^{2N-1}\Pi_m),
\end{equation}
\begin{equation}
\overline{\Pi}^{(2N)}_{o} = \sum_{m=0}^{N-1}\Pi_{2m+1} + \frac{1}{2}(I-\sum_{m=0}^{2N-1}\Pi_m).
\end{equation}
Thus, our POVM is $\{\overline{\Pi}^{(2N)}_{e},\overline{\Pi}^{(2N)}_{o}\}$. The corresponding GYNI inequality and its violation can then be studied in terms of functions of $|\alpha|^2$, $N$, and $\kappa$.
In order to compare with the effect of lossy detector on the single photon entangled state, we use the same POVM but label the outcomes differently: $\{a=0{\equiv}\overline{\Pi}^{(2N)}_{o},a=1{\equiv}\overline{\Pi}^{(2N)}_{e};b=0{\equiv}\overline{\Pi}^{(2N)}_{e},b=1{\equiv}\overline{\Pi}^{(2N)}_{o}\}$. In case of the single-photon state the left hand side of the GYNI inequality is simply $\kappa$ (irrespective of $N$), similar to what is observed in the case of the lossy beam splitter.
\subsubsection*{Case 2: When the detector can resolve upto $2N-1$ photons}
\begin{figure}[ht]
	\centering
	\includegraphics[width=12cm, height=6cm]{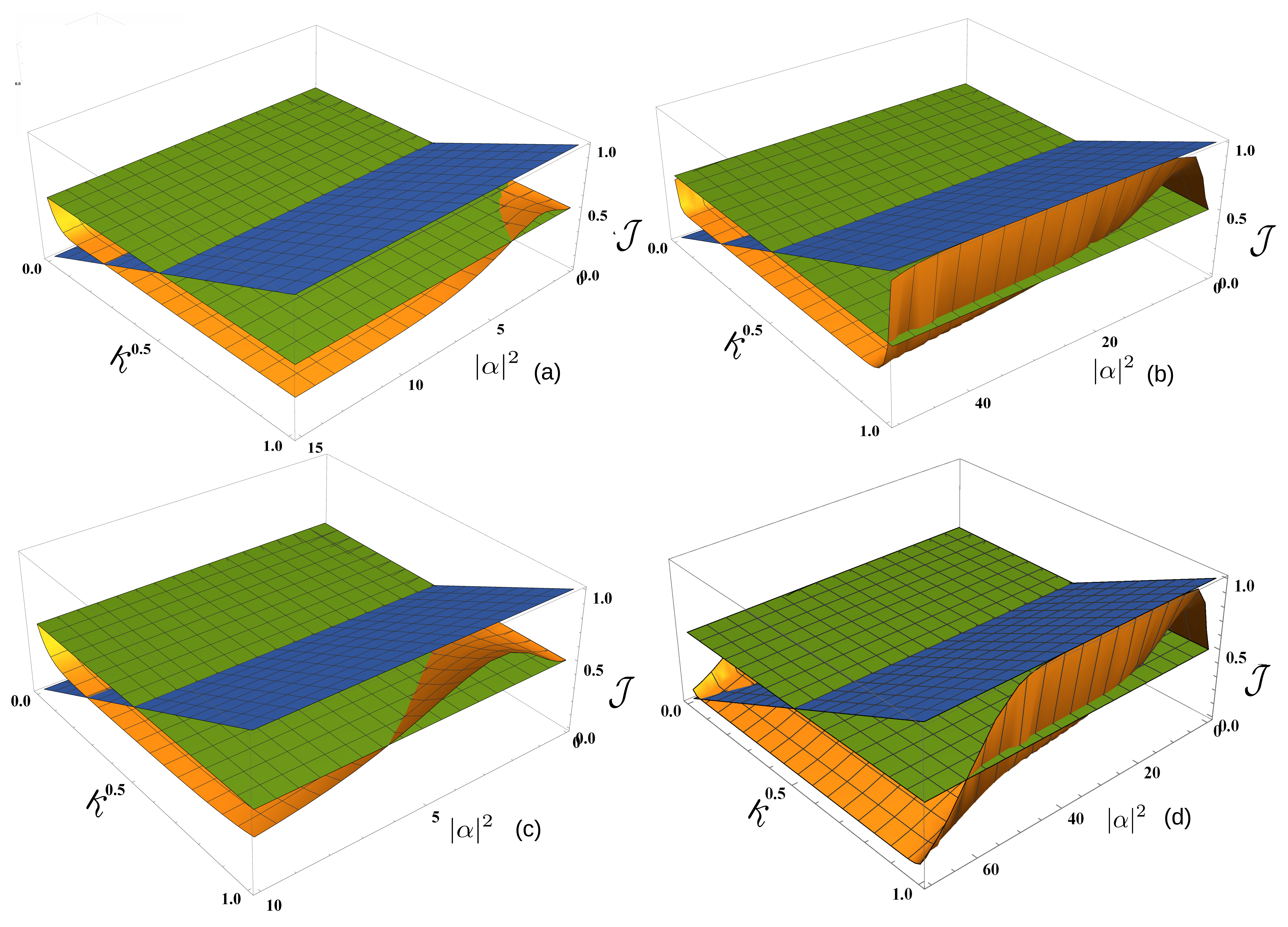}
	\caption{(Colour online) The green planes $z=0.5$ mark the classical bound, i.e. the threshold (classical bound of 0.5). The blue planes given by the simple function $\kappa$, mark the GYNI violation for the single photon state. The yellow surfaces mark the functions $\mathcal{J}$ (inequality (\ref{eq:gyni}) furnished by the even-coherent NOON state. \textbf{(a)} $N=1$; POVM =$\{\overline{\Pi}^{(2N)}_{e},\overline{\Pi}^{(2N)}_{o}\}$; violation is seen when $\kappa$ is close to 1 and when the average photon number is small and the curve drops below the threshold value as the average photon number is inclreased;  \textbf{(b)} $N=20$; POVM =$\{\overline{\Pi}^{(2N)}_{e},\overline{\Pi}^{(2N)}_{o}\}$; maximal violation is observed for large values of average photon number and this persists for a large range of $|\alpha|^2$ before eventually dropping; \textbf{(c)} $N=1$; POVM =$\{\overline{\Pi}^{(2N-1)}_{e},\overline{\Pi}^{(2N-1)}_{o}\}$; violation is observed when $\kappa$ is close to unity and $|\alpha|^2$ is small; there is a drop in violation when $|\alpha|^2$ increases; \textbf{(d)} $N=20$; POVM=$\{\overline{\Pi}^{(2N-1)}_{e},\overline{\Pi}^{(2N-1)}_{o}\}$;maximal violation is observed (and is seen to persist) when $\kappa$ is close to unity and $|\alpha|^2$ is large.}
	\label{fig:fig3}
\end{figure}

The structure of the POVM in this case follows almost immediately from the previous one:
\begin{equation}
\overline{\Pi}^{(2N-1)}_{e} = \sum_{m=0}^{N-1}\Pi_{2m} + \frac{1}{2}(I-\sum_{m=0}^{2N-2}\Pi_m),
\end{equation}
\begin{equation}
\overline{\Pi}^{(2N-1)}_{o} = \sum_{m=0}^{N-2}\Pi_{2m+1} + \frac{1}{2}(I-\sum_{m=0}^{2N-2}\Pi_m).
\end{equation}
The POVM is $\{\overline{\Pi}^{(2N-1)}_{e},\overline{\Pi}^{(2N-1)}_{o}\}$.
The corresponding GYNI inequality and its violation can, once again, be studied in terms of functions of $|\alpha|^2$, $N$, and $\kappa$. Once again, in order to compare the effect of the lossy detector with odd saturation on the single-photon entangled state with that on the even-coherent NOON state, the same POVM but label the outcomes differently: $\{a=0{\equiv}\overline{\Pi}^{(2N-1)}_{o},a=1{\equiv}\overline{\Pi}^{(2N-1)}_{e};b=0{\equiv}\overline{\Pi}^{(2N-1)}_{e},b=1{\equiv}\overline{\Pi}^{(2N-1)}_{o}\}$. Using this POVM on the single-photon state  the left hand side of the GYNI inequality is once again found to be $\kappa$, irrespective of $N$. The results are summarized in Fig. (\ref{fig:fig3}). It is to be noted that the violation of the GYNI inequality furnished by the even-coherent NOON state persists when the efficiency $\kappa$ is close to unity, even if the saturation number and average photon number are small. For larger values of saturation, the even-coherent NOON state furnishes maximum violation of the GYNI inequality, in a certain range of $|\alpha|^2$.

%%%%%%%%%%%%%%%%%%%%%%%%%%%%%%%%%%%%%%%%%%%%%%%%%%%%%%%%%%%5
\section{Bell correlators}
\label{ap:bell}
 As mentioned in section \ref{sec:nonloc}, the Bell correlators are found by performing displacement measurements on each of the two modes of the two-mode entangled states.
 \subsection{Even-coherent NOON state}
 \begin{align}
 \expval{M^A_0 M^B_0} & = N [6 e^{-|\alpha|^2} - (1+ e^{-2|\alpha|^2})], \\
 \expval{M^A_0 M^B_1} & = N [2 (4  e^{-|\alpha|^2}  - (1+ e^{-2|\alpha|^2})) (2 e^{-|\beta_2|^2} -1) 
 +  2 (e^{-|\alpha + \beta_2|^2 } +  e^{-|\alpha - \beta_2|^2 } \nonumber \\ 
 & + 2 e^{-(|\alpha|^2 + |\beta_2|^2 )} \cos{(2\Im (\alpha \beta_2^\ast))} - (1 + e^{-2|\alpha|^2})) \nonumber\\ 
 &+ 4 e^{-|\alpha|^2} (2 e^{|\beta_2|^2} (\cosh{(\alpha \beta_2^\ast)} + \cosh{(\alpha^\ast \beta_2) -1}], \\
 \expval{M^A_1 M^B_0} & = N [2 (4  e^{-|\alpha|^2}  - (1+ e^{-2|\alpha|^2})) (2 e^{-|\beta_1|^2} -1) 
 +  2 (e^{-|\alpha + \beta_1|^2 } +  e^{-|\alpha - \beta_1|^2 } \nonumber \\ 
 & + 2 e^{-(|\alpha|^2 + |\beta_1|^2 )} \cos{(2\Im (\alpha \beta_1^\ast))} - (1 + e^{-2|\alpha|^2})) \nonumber\\
 & + 4 e^{-|\alpha|^2} (2 e^{|\beta_1|^2} (\cosh{(\alpha \beta_1^\ast)} + \cosh{(\alpha^\ast \beta_1) -1}],  
 \end{align}
 and
\begin{align}
 \expval{M^A_1 M^B_1} & = N [ 2(e^{-|\beta_1|^2} - 1 ) \{ e^{-|\alpha + \beta_1|^2 } +  e^{-|\alpha - \beta_1|^2 }\nonumber\\
 & + 2 e^{-(|\alpha|^2 + |\beta_1|^2 )} \cos{(2\Im (\alpha \beta_1^\ast))} - (1 + e^{-2|\alpha|^2})\} \nonumber \\ 
 & +  2(e^{-|\beta_2|^2} - 1 ) ( e^{-|\alpha + \beta_2|^2 } +  e^{-|\alpha - \beta_2|^2 }  \nonumber\\
 &+ 2 e^{-(|\alpha|^2 + |\beta_2|^2 )} \cos{(2\Im (\alpha \beta_2^\ast))} - (1 + e^{-2|\alpha|^2}) \nonumber \\
 & + 4 e^{-|\alpha|^2} [(2e^{-|\beta_1|^2} \cosh{(\alpha^\ast \beta_1)} -1) (2e^{-|\beta_2|^2} \cosh{(\alpha \beta_2^\ast)} -1) \nonumber \\
 & + (2e^{-|\beta_1|^2} \cosh{(\alpha \beta_1^\ast)} -1) (2e^{-|\beta_2|^2} \cosh{(\alpha^\ast \beta_2)} -1) ]],
 \end{align}
 where $N = \frac{1}{(1+e^{-|\alpha|^2})^2}$.
\subsection{Coherent NOON state} 
\begin{align}
\expval{M^A_0 M^B_0} & = N^{\prime}(3e^{-|\alpha|^2}-1),\\
\expval{M^A_0 M^B_1} & = N^{\prime}(2e^{-(|\alpha|^2+r^2)}+e^{-|\alpha-\beta_2|^2}+2e^{\frac{-(|\alpha|^2+r^2)}{2}}e^{\frac{-|\alpha-\beta_2|^2}{2}}\cos(Im\{\alpha\beta^*_2\})\nonumber\\
&-e^{-r^2}-2e^{-|\alpha|^2}),\\
\expval{M^A_1 M^B_0} & = N^{\prime}(2e^{-(|\alpha|^2+r^2)}+e^{-|\alpha-\beta_1|^2}+2e^{\frac{-(|\alpha|^2+r^2)}{2}}e^{\frac{-|\alpha-\beta_1|^2}{2}}\cos(Im\{\alpha\beta^*_1\})\nonumber\\
&-e^{-r^2}-2e^{-|\alpha|^2}),\\
\text{and}\hspace{1mm}\expval{M^A_1 M^B_1} & =
N^{\prime}(2(e^{-(abs(a-bb1))**2}e^{-r^2}+e^{-|\alpha-\beta_2|^2}e^{-r^2}\nonumber\\
&+e^{\frac{-(r^2+|\alpha|^2)}{2}}e^{-\frac{|\alpha-\beta_2|^2+|\alpha-\beta_2|^2}{2}}2\cos(Im\{\alpha\beta^*_1+\alpha^*\beta_2\}))\nonumber\\
&-(e^{-r^2}+e^{-|\alpha-\beta_2|^2}+2e^{\frac{-(|\alpha|^2+r^2)}{2}}e^{-(|\alpha-\beta_2|^2)/2}\cos(Im\{\alpha\beta^*_2\}))\nonumber\\
&-(e^{-r^2}+e^{-|\alpha-\beta_1|^2}+2e^{\frac{-(|\alpha|^2+r^2)}{2}}e^{-|\alpha-\beta_1|^2/2}\cos(Im\{\alpha\beta^*_2\})))+1,
\end{align}
 where $N^{\prime} = \frac{1}{1+e^{-|\alpha|^2}}$.
\subsection{Photon-added coherent NOON state}
\label{ap:subsec_pa}
\begin{align}
\expval{M^A_0 M^B_0} & = -1,\\
\expval{M^A_0 M^B_1} & = \frac{e^{-(|\alpha|^2+|\beta_2|^2)}}{1+|\alpha|^2}[|\beta_2|^2e^{2Re\{\alpha^*\beta_2\}}]-e^{-|\beta_2|^2},\label{eq:dis1}\\
\expval{M^A_1 M^B_0} & = \frac{e^{-(|\alpha|^2+|\beta_1|^2)}}{1+|\alpha|^2}[|\beta_1|^2e^{2Re\{\alpha^*\beta_1\}}]-e^{-|\beta_1|^2},\label{eq:dis2}\\
\expval{M^A_1 M^B_1} & =
1 + \frac{2e^{-|\alpha|^2}}{1+|\alpha|^2}[|\beta_1|^2e^{-|\beta_2|^2+2Re\{\alpha^*\beta_1\}}+|\beta_2|^2e^{-|\beta_1|^2+2Re\{\alpha^*\beta_2\}} \nonumber\\
&+e^{\frac{-(|\beta_1|^2+|\beta_2|^2)}{2}}\{\beta_1\beta^*_2e^{\alpha*\beta_1+\alpha\beta*_2}+\beta_2\beta^*_1e^{\alpha*\beta_2+\alpha\beta*_1}\}] -\expval{M^A_0 M^B_1} -\expval{M^A_1 M^B_0},\label{eq:dis3}
\end{align}
where the last two terms in eq. (\ref{eq:dis3}) are given by eq.s (\ref{eq:dis1}) and (\ref{eq:dis2}).
%%%%%%%%%%%%%%%%%%%%%%% References %%%%%%%%%%%%%%%%%%%%%%%%%
\bibliography{research,thesis}

\begin{thebibliography}{62}
\bibitem [{gen()}]{gen_r}%
  \BibitemOpen
  \href@noop {} {}\bibinfo {note} {The state
  $\ket{\xi_1}\ket{\xi_2}+\ket{\xi_2}\ket{\xi_1}$, with
  $\ket{\xi_i}\hspace{1mm}(i{\in}\{1,2\})$ being nonclassical states, also
  turns out to be advantageous in two-way communication and may have an average
  photon number lower than unity. However, the use of these states defeats our
  purpose of using a single-information carrier for two-way
  communication}\BibitemShut {NoStop}%
\end{thebibliography}

\end{document}